\documentclass[aip,floatfix,amsmath,amssymb,reprint]{revtex4-1}
\pdfoutput=1
\usepackage{graphicx}% Include figure files
\usepackage{dcolumn}% Align table columns on decimal point
\usepackage{bm}% bold math
\usepackage[utf8]{inputenc}
\usepackage[T1]{fontenc}
\usepackage{mathptmx}
\usepackage{etoolbox}
\usepackage{subfigure}
\usepackage{color,soul}
\usepackage{comment}
\usepackage{orcidlink}
\usepackage{placeins}

\draft % marks overfull lines with a black rule on the right

\begin{document}

%\preprint{}

\title{Instability of a Counter-Streaming Collisionless Pair Plasma II: Angular and Spectral Analysis of the Weibel Instability} %Title of paper



\author{Michael C. Sitarz \orcidlink{0000-0001-9003-0737}}
\affiliation{Department of Physics and Astronomy, University of Kansas, Lawrence, KS 66045}
\email{mcsitarz@ku.edu}

\date{April 23, 2024}

\begin{abstract}
Energetic astrophysical phenomena, such as $\gamma$-ray bursts, supernova explosions, and magnetar flares occur in collisionless plasmas and involve various plasma kinetic and magnetohydrodynamic instabilities. In this paper, we explore the spectral trends of the Weibel instability using spectral analysis of particle-in-cell simulations. Power dependence on viewing angle and frequency are explored and the relation to the results of the first paper in this series is discussed.
\end{abstract}


\maketitle %\maketitle must follow title, authors, abstract and \pacs

\section{Introduction}
\indent Supernova blast waves (SN) \cite{Guo}, interplanetary medium (IPM) shocks \cite{2}, quasar jets \cite{2,3}, solar flares \cite{4}, pulsar wind nebula (PWN) \cite{5}, relativistic jets in active galactic nuclei (AGN), gamma ray bursts (GRB) \cite{6}, and the virilization of intergalactic medium (IGM) all exist in a plasma like environments that are energetic enough to generate collective phenomena. Often, examining the wave modes such a system produces can illuminate the processes and mechanisms within the system. \\

\indent Weibel \cite{7} and Fried \cite{8} demonstrated that magnetic fields arise from an instability driven by particle distribution function (PDF) anisotropy. The anisotropy present in either velocity- or temperature-space brings forth a plasma instability responsible for runaway magnetic fields. One of the more common set ups for this system is used in this series of papers - the counter streaming beams - studying specifically the counter streaming beams of a $e^\pm$ pair plasma. While much focus has been spent on studying the radiation of the Weibel system (for example, Jitter radiation produced as a by-product of particle acceleration during filament building \cite{9}), there is still many questions surrounding the spectral signals of such a system. This paper puts forth an analysis of the complex spectral space generated by two counter streaming pair plasma beams. \\

\indent In the previous paper of this series \cite{MCS_WP1}, hereafter referred to as Paper I, it was put forth that different system parameters (like beam propagation speed) have corresponding outcomes on the physical evolution and field generation of the Weibel instability (WI) system . These outcomes dictate what type of evolution the system as a whole will have, or if the WI will evolve toward non-linearity and turbulence as opposed to thermalization. Within these physical system exists a spectral system with similar outcomes. Paper I\cite{MCS_WP1} briefly explored the spectral regime, identifying the dominate mode to be an isotropic electromagnetic (EM) wave, visible in both magnetic and electric fields. But other wave modes were identified in the $E_x$ and $E_y$ fields separate to the dominant mode. These wave modes can be brought about by the two stream instability (TSI) working in tandem and after the WI.\\

\indent This paper presents a study on the different electrostatic spectral wave modes found in the WI/TSI system. This study and conclusions are a result of first principle simulations and spectral signatures. The Weibel instability generated by two counter streaming beams in time from the ``cold beam''  system evolves in time into a ``warm plasma'' distribution, where the momenta can be described by Maxwell-J{\"u}ttner Distribution \cite{10}. It is at the critical point of saturation of the filaments where the Weibel instability approaches a breakdown. As the plasma transforms into the ``warm” regime through the dissipation and merger of the saturated filaments, the particles interact with the dominant electromagnetic (EM) mode and a possible secondary instability (TSI) is generated within the system. \\

\indent The remainder of the paper will be organized as follows: \S$2$ is a review of the Weibel and Two Stream instabilities, \S$3$ is a description of the simulation set up and a discussion of the analytical techniques used in the study of the data, and \S$4$ will discuss the results of the analysis and the conclusions drawn from the study. Finally, \S$5$ will contain concluding remarks and possible implications.\\

\section{Major Instability Background}
\subsection{Weibel Instability}
\indent A shortened version of the WI timeline, filament theory, and dispersion relation taken from Paper I \cite{MCS_WP1} follows below. For the full treatment of the WI, please see the previous paper.

\subsubsection{Weibel Instability Timeline}
\indent The Weibel instability (WI) is claimed to be the source of intense magnetic fields within the GRB prompt emission, afterglow \cite{9} and astrophysical shock frames \cite{11}, which were later proved numerically with strong collaborating evidence provided by numerous numerical simulations by a number of sources \cite{12,13,14,15,16}. The GRB shock is mediated by the mechanisms generated by the WI, located at the shock front \cite{19}. This turbulence is sub-Larmor in scale \cite{18} and of a few ion skin depths \cite{17}. Small scale turbulence like this is common in various astrophysical settings, some examples include Whistler filamentation and mixed modes or electrostatic Langmuir oscillations \cite{9}. \\

\indent In $1959$, Weibel considered a non-relativistic plasma with an anisotropic particle distribution function (PDF), analyzing it with a fully kinetic analysis \cite{7}. Later that same year, Fried \cite{8} treated this same PDF more generally as a cold plasma with two counter streaming beams. These beams, comprised of electrons, were perpendicularly threaded by a small sinusoidal magnetic field. This magnetic field goes on to produce a runaway effect, creating the plasma instability. The runaway effect is a temporary unstoppable cycle of field deflecting particles, strengthening current densities, deflecting more particles, strengthening fields, etc. \\

\indent In linear plasma wave analysis analysis, this sinusoidal field signature can be treated as the initial fluctuations seen in the system. As a consequence of this field, and the Lorentz force ($\vec{F} = \frac{e}{c}(\vec{v} \times \vec{B})$), charged particles are deflected from their beam trajectories into concentrated nodes in the system. These nodes then produce current filaments, which are responsible for the salient tiger striped feature of the instability. The magnetic field in the filaments then increase the magnitude of the initial fluctuations. As the system continues this runaway effect, the growth rate of the instability reaches a maximum of 
\begin{equation}\label{Eq:Max_Growth_Rate}
    \gamma_{max} \approx \gamma^{\frac{1}{2}}\omega_p \rightarrow \frac{c}{\lambda_{De}},
\end{equation}
with the fastest growing mode defined as
\begin{equation}\label{Eq:Fastest_Mode}
    k_B = \frac{\omega_{p,s}}{c}.
\end{equation}
From the above expressions, the plasma frequency $\omega_p = \sqrt{\frac{4\pi e^2}{m}}$, and Debye length $\lambda_{De} = \frac{v_{th}}{\omega_p} = \sqrt{\frac{kT_e}{4\pi ne^2}}$ set the correlation scale of the produced magnetic fields. \\

\indent Particles are continually deflected in larger and larger quantities, which in turn amplifies the magnetic fields. Deflections of particle orbits in the linear scale are seen on the scale of the Larmor radius $\rho_L = v_{\perp,B}/\omega_{c,s}$ (with cyclotron frequency $\omega_{c,s} = \frac{eB}{m_sc}$). These deflections continually increase as the field increases. Particle motion, at this stage, is limited to along the field, with free streaming suppressed by these deflections. When the ever growing magnetic field reaches $k_b\rho_L \sim 1$, particles can no longer be deflected into nodes and are trapped, halting the cyclical amplification and saturating the instability. This is when the system begins to breakdown. As the breakdown of the runaway affect begins, this is where the evolution of the WI splits between thermalization and the TSI.

\subsubsection{Current Filament Theory}
\indent This discussion is derived from Medvedev et. al. \cite{20} and the sources therein. Without loss of generality, filamentation theory is identical in higher dimensions, therefore, this discussion will consider only one-dimensional filaments. At the start of the instability, all filaments form identically with diameter $D_0$, mass per unit length $\mu_0 \simeq 0.25mnD_0^2\pi$ (particle mass $m$ and number density $n$), current $I_0$, and spatial separation $d_0 \simeq 2D_0$ from center to center. There is no pattern to filament placement in the system, they are randomly distributed throughout the system and begin at rest, but remain vulnerable to attraction/repulsion forces.\\

\indent As more particles are deflected into nodes, filaments begin to grow more and more, which in turn amplify the field. They grow in transverse scale beginning with the skin length scale, $\lambda_B \approx 2\pi c/\omega_{p,e^-}$, and continue to scale with the correlation length of the produced fields \cite{20}. Filament growth rate decreases as temperature increases, but is not constrained \cite{21}, allowing filaments to grow unopposed parallel to beam direction. When the instability reaches saturation and the magnetic field begins to decrease, currents begin to drift toward other like-current filaments (as taught by the right hand rule) where the filaments begin to orientate into a more regular pattern. As they drift toward each other, some begin to merge and coalesce. \\

\indent The instant two filaments touch is considered a filament merger, as opposed to a full integration into a single filament. This occurs when $d_0 \simeq D_0$. The overall merger rate is limited as filaments coalesce and become further away from other mergers, as the force interaction weakens as distance increases. Merging is a self-similar and hierarchical process, with $N_0$ number of filaments merging pairwise into a new ($1^{st}$) generation of filaments numbering $N_0/2$. The properties of the initial filaments scale as the $k^{th}$ generation is produced, following
\begin{multline}\label{Eq:Number_of_kth_Gen_Fil}
    I_k = 2^kI_0, \ \mu_k = 2^k\mu_0, \ D_k = 2^{k/2}D_0, \\
    d_k \sim \frac{D_k}{2}, \ \tau_{k,NR} = \tau_{0,NR}, \\
    \tau_{k,R} = 2^{k/2}\tau_{0,R}.
\end{multline}

\subsubsection{Analytical Dispersion Relation}
\indent The following derivation is a mix of sources from \cite{22,23,24}, we encourage the reader to consult these papers for a more detailed review. \\

\indent The initial set-up contains a uniform plasma with immobile and neutralizing ion background and two counter streaming $e^-$ beams. The ion background has density $n_i = \sum_\iota n_{o,\iota}$ over $\iota$ species of ion. It is assumed that for all ions $Z = 1$. For this derivation, ``ion'' is simply a particle that is not an electron. The beams will propagate in a $2D$ box along the $\hat{x}$ direction with unperturbed number density $n_{0,\alpha}$, with $\alpha$ denoting the electron species of either positron on electron (beam population). Together with their velocities $v_{0,x,\alpha}$, the beams contribute zero global current density
\begin{equation}\label{Eq:Beam_Current_Density}
    \sum_\alpha n_{0,\alpha} v_{beam,x,\alpha} = 0.
\end{equation}
The three-momenta of the beams can be further defined by
\begin{equation}\label{Eq:Beam_Three_Momenta}
    \vec{v}_\alpha = \frac{\vec{p}_\alpha c}{\sqrt{m^2c^2 + p_\alpha^2}}.
\end{equation}

\indent Begin with Maxwell's equations
\begin{equation}\label{Eq:Gauss_Law_Max}
    \nabla \cdot \vec{E} = \sum_\alpha n_{beam,\alpha},
\end{equation}
\begin{equation}\label{Eq:No_Mono_Max}
    \nabla \cdot \vec{B} = 0,
\end{equation}
\begin{equation}\label{Eq:Faraday_Law_Max}
    \nabla \times \vec{E} = \frac{-\partial B}{\partial t},
\end{equation}
and
\begin{equation}\label{Eq:Ampere_Law_Max}
    \nabla \times \vec{B} = \sum_\alpha n_{beam,\alpha} \vec{v}_{beam,x,\alpha} + \frac{1}{c^2}\frac{\partial \vec{E}}{\partial t},
\end{equation}
the relativistic dynamics of $e^-$
\begin{equation}\label{Eq:Rel_Momenta_Electron_MHD}
    \frac{\partial \vec{p}_\alpha}{\partial t} + (\vec{v}_\alpha \cdot \nabla)\vec{p}_\alpha = -e\left(\vec{E} + \frac{\vec{v}_\alpha}{c} \times \vec{B}\right),
\end{equation}
\begin{equation}\label{Eq:Continuity_Electron_MHD}
    \frac{\partial n_{beam,\alpha}}{\partial t} + \nabla \cdot n_{0,\alpha}\vec{v}_\alpha = 0,
\end{equation}
and vector potential
\begin{equation}\label{Eq:Mag_Vector_Potential}
    \vec{B} = \nabla \times \vec{A},
\end{equation}
\begin{equation}\label{Eq:Momenta_due_to_Vector_Potential}
    p_{l,\alpha} - \frac{eA_l}{c} = p_{beam,l,\alpha},
\end{equation}
where $l$ represents a coordinate ($x$, $y$, $z$). The following substitution has been made for explicitness 
\begin{equation}\label{Eq:Current_Density_Beam}
    \vec{j} = -n_{beam,\alpha}\vec{v}_{beam,x,\alpha}.
\end{equation}

\indent Coupling the $e^-$ momentum and density with the Maxwell equations recovers the following without loss of generality with coordinates ($l$, $m$, $n$)
\begin{equation}\label{Eq:Max_Gauss_Part}
    \frac{\partial E_l}{\partial l} = 4\pi e\left(n_i - \sum_\alpha n_\alpha\right),
\end{equation}
\begin{equation}\label{Eq:Max_Ampere_Part}
    \frac{\partial B_n}{\partial m} = -\frac{4 \pi e}{c}\sum_\alpha n_\alpha v_{x,\alpha} + \frac{1}{c}\frac{\partial E_l}{\partial t},
\end{equation}
\begin{equation}\label{Eq:Max_Double_Faraday_Part}
    -\frac{\partial^2 E_l}{\partial m^2} = -\frac{1}{c}\frac{\partial}{\partial t}\frac{\partial B_n}{\partial m},
\end{equation}
\begin{equation}\label{Eq:Max_Faraday_Part}
    \frac{1}{c}\frac{\partial B_n}{\partial t} = \frac{\partial E_l}{\partial m},
\end{equation}
\begin{equation}\label{Eq:Momenta_Part}
    \frac{\partial p_{l,\alpha}}{\partial t} + v_{m,\alpha}\frac{\partial p_{m,\alpha}}{\partial m} = -e \left(E_m - \frac{v_{x,\alpha}}{c}B_n \right),
\end{equation}
\begin{equation}\label{Eq:Continuity_Part}
    \frac{\partial n_\alpha}{\partial t} + \frac{\partial n_\alpha v_{m,\alpha}n_{m,\alpha}}{\partial m} = 0.
\end{equation}

\indent The above equations are now linearized using a small plane wave perturbation of the form 
\begin{equation}\label{Eq:Linear_Plane_Wave}
    F(x,y,t) = fexp[ik_xx - ik_yy - i\omega t],
\end{equation}
applied to the velocities, densities, and fields. $\omega$ represents the angular frequency perturbation while $k_i$ represents the wave-vector perturbation. In the linearization calculations, the following relationships are used: $\frac{\partial}{\partial t} \rightarrow -i\omega$ and $\nabla \rightarrow ik_y - ik_x$.\\

\indent For the remainder of this example, we will only be concerned with variables with respect to $y$ and $t$. This truncation of dimensions is valid due to the nature of the WI and the Two Stream instability (TSI). $k_y = 0$, from this plane wave perturbation process, gives the TSI. If there is oblique and intermediate propagation angles where $k_x$ and $k_y$ are non-vanishing, the WI and the TSI are coupled into a single branch. In one dimension, the TSI has a set cutoff at $k_x^{Max}$ beyond which the TSI is no longer unstable. The full, $3$-space dispersion can be found in the mentioned sources and will be examined in depth in subsequent publications.\\

\indent Linearizing (Eqs. \ref{Eq:Max_Gauss_Part}, \ref{Eq:Max_Ampere_Part}, \ref{Eq:Max_Double_Faraday_Part}, \ref{Eq:Max_Faraday_Part}, \ref{Eq:Momenta_Part}, \ref{Eq:Continuity_Part}) and substituting them into (Eq. \ref{Eq:Max_Faraday_Part}) recovers a sixth order relation for the WI
\begin{equation}\label{Eq:6th_Order_DR}
    (\omega^2 - \Omega_a^{2})[\omega^4 - \omega^2(k^2c^2 + \Omega_b^2) - k^2c^2\Omega_c^2] - k^2c^2\Omega_d^2 = 0.
\end{equation}
Here, the following substitutions are used 
\begin{equation}\label{Eq:DR_Omega_a}
    \Omega_a^2 = \omega_{pe}^2 \sum_\alpha \frac{n_{beam,\alpha}}{n_i \Gamma_\alpha},
\end{equation}
\begin{equation}\label{Eq:DR_Omega_b}
    \Omega_b^2 = \omega_{pe}^2 \sum_\alpha \frac{n_{beam,\alpha}}{n_i \Gamma_\alpha^3},
\end{equation}
\begin{equation}\label{Eq:DR_Omega_c}
    \Omega_c^2 = \omega_{pe}^2 \sum_\alpha \frac{n_{beam,\alpha} v_{beam,x,\alpha}^2}{n_i \Gamma_\alpha c^2},
\end{equation}
\begin{equation}\label{Eq:DR_Omega_d}
    \Omega_d^2 = \omega_{pe}^2 \sum_\alpha \frac{n_{beam,\alpha} v_{beam,x,\alpha}}{n_i \Gamma_\alpha c},
\end{equation}
where
\begin{equation}\label{Eq:Lorentz_Factor_Beam}
    \Gamma_\alpha = \left(1 - \frac{v_{beam,x,\alpha}^2}{c^2}\right)^{-1/2}.
\end{equation}
For a $e^\pm$ beam, we can further denote the beam density as 
\begin{equation}\label{Eq:Beam_Density_DR}
    n_{beam, e^-} = n_{beam, e^+} = n_{beam}/2,
\end{equation}
with velocities
\begin{equation}\label{Eq:Beam_Velocity_DR}
    v_{beam, e^-} = v_{beam, e^+} = v_{beam,x}/2.
\end{equation}
We may also remove the ion background $n_i$ without disrupting the following relations, as it was immobile and present for charge neutrality. 

\indent This dispersion relation can be solved by using substitution ($u = \omega^2$), which reduces the order of the function, and then employing the companion matrix method to find the eigenvalues (roots). Three branches can be found from the eigenvalues of the function. Two real ($\mathbb{C} = 0$), oscillatory modes and a single exponentially growing mode with imaginary components ($\mathbb{C} \ne 0$). This exponentially growing mode is the WI. The maximum growth rate $\gamma_{Max}$ is found in the short wave limit where $k^2c^2$ dominates over the $\Omega_i$ terms. This $\gamma$ is not the same as $\Gamma_\alpha$, which is used for kinematics. When the growth rate is discussed, the $\gamma$ will be explicitly mentioned to avoid confusion. Applying this condition to (Eq. \ref{Eq:6th_Order_DR}) finds
\begin{equation}\label{Eq:WI_Growth_DR}
    \gamma_{Growth \ Rate} \approx \frac{\sqrt{\sqrt{(\Omega_a^2 + \Omega_c^2)^2 - 4\Omega_d^4} - (\Omega_a^2 - \Omega^2_c)}}{\sqrt{2}}.
\end{equation}
Long wave dependence ($k^2c^2 \sim 0$) gives
\begin{equation}\label{Eq:WI_Growth_DR_LongWave}
    \gamma_{Growth \ Rate} \approx \sqrt{\frac{\Omega_a^2\Omega_c^2 - \Omega_d^4}{\Omega_a^2\Omega_b^2}}.
\end{equation}

\subsection{Two Stream Instability}
\subsubsection{Instability Origins}
\indent The following discussion and derivation is inspired by notes taken from the lectures of Ling-Hsiao Lyu \cite{TSI-Notes}. The authors encourage the reader to read the sources for more detailed information. \\

\indent The two stream instability (TSI) is one of the most ubiquitous instabilities found in plasma physics. It occurs when two species of particles have counter propagating drift velocities $v_0$. From this basic physical set-up, further system parameters can dictate the unstable electrostatic modes within the dispersion relation. \\

\indent There are two general beam cases when considering the TSI, cold and hot, that both saturate when the beam particles are bound within the electric field of the propagating wave. For hot beams, TSI can be thought of as a type of inverse Landau damping. There is a small population of particles with a drift velocity greater than that of the phase velocity of the propagating wave. The majority of particles are slower than the phase velocity and there are particles with equal velocities to the wave. Regarding the instability system itself, when a hot beam of electrons is injected into a stationary background, the velocity space distribution is said to posses a ``bump on tail'' distribution function. On this function, if the phase velocity of the excited wave exists in a region of positive slope (more particles faster than its phase velocity than slower) there exists a greater energy transfer from the fast particles to the slower wave, further exciting the wave. For cold beams, none of the particles in either beam possess a drift velocity equal to that of the phase velocity of the wave within the system (resonance). The beam particles are clustered in physical space in a propagating wave. This motion becomes self-reinforcing despite no resonance. The excitation of a TSI in the cold beam limit is discussed below.\\

\subsubsection{\label{Sec:TSI_DR}Analytical Dispersion Relation}
\indent A general dispersion relation for the TSI is seen below, a more specific dispersion relation derivation pertaining to the system present in this study and its complex field dynamics will be done in later publications. For this example, we will be setting the magnetic field, $\vec{B}$, to zero. This will allow only electrostatic modes to be present in the derivation. In many classical examples, there is a neutralizing, frozen in space ion background. The ion background may be treated as fixed because of the separation in relevant dynamical scales between ions and electrons on account of their disparate masses. For electron-positron plasma, both species are equal and active participants in the system. Without loss of generality or accuracy, this example will use electrons and positrons ($m_{e^-} = m_{e^+}$). This will conserve current density and charge neutrality the same as mobile electrons and immobile ions. Nevertheless, each system consists of two cold species (subscript $1$, $2$, or $s$) with initial constant drift velocity $v_{0,1}$ and $v_{0,2}$. This gives a particle momentum equation for each species of 
\begin{equation}\label{Eq:Particle_Momenta_Equation}
    \frac{\partial \vec{v_{0,s}}}{\partial t} + (\vec{v_{0,s}} \cdot \nabla)\vec{v_{0,s}} = \frac{q_s}{m_s}\vec{E},
\end{equation}
and the continuity equation for each species
\begin{equation}\label{Eq:Particle_Continuity_Equation}
    \frac{\partial n_{0,s}}{\partial t} + \nabla \cdot n_{0,s}\vec{v_{0,s}} = 0.
\end{equation}
Because we do not worry about $\vec{B}$ in this example, the only Maxwell equation we will need is Gauss's equation
\begin{equation}\label{Eq:Maxwell_Gauss_Equation}
    \nabla \cdot \vec{E_0} = 4 \pi \sum_s q_s n_{0,s}.
\end{equation}

\indent Replacing the initial values for electric field, drift velocity, and particle density with first order perturbation summations of the form (see also the plane wave perturbation used in (Eq. \ref{Eq:Linear_Plane_Wave}))
\begin{equation}\label{Eq:Peerturbation_Form}
    \tilde{E} = \vec{E_0} + \vec{E_1}.
\end{equation}
Removing any terms of order zero or order two, the linearized expressions of equations (Eq. \ref{Eq:Particle_Momenta_Equation}, Eq. \ref{Eq:Particle_Continuity_Equation}, and Eq. \ref{Eq:Maxwell_Gauss_Equation}) now read
\begin{equation}\label{Eq:Linearized_Momenta_Equation}
    \frac{\partial \tilde{v_s}}{\partial t} + \vec{v_{0,s}} \cdot \nabla \tilde{v_s} = \frac{q_s}{m_s} \tilde{E},
\end{equation}
\begin{equation}\label{Eq:Linearized_Continuity_Equation}
    \frac{\partial \tilde{n_s}}{\partial t} + n_{0,s}\nabla \cdot \tilde{v_s} + \vec{v_{0,s}} \cdot \nabla \tilde{n_s} = 0,
\end{equation}
\begin{equation}\label{Eq:Linearized_Gauss_Equation}
    \nabla \cdot \tilde{E} = 4\pi \sum_s q_s \tilde{n_s}.
\end{equation}

\indent We may now replace the terms of the form $\tilde{E}$ with the plane wave perturbation expression
\begin{equation}\label{Eq:Plane_Wave_Perturbation_Equation}
    \tilde{E} = \vec{E_0}e^{i(\vec{k}\cdot\vec{r} - \omega t)}.
\end{equation}
This transforms the linearized momenta equation (Eq. \ref{Eq:Linearized_Momenta_Equation}) to
\begin{equation}\label{Eq:Fourier_Momenta_Equation}
    (-i\omega + i\vec{k} \cdot \vec{v_{0,s}})\tilde{v_s} = \frac{q_s}{m_s} \tilde{E} ,
\end{equation}
 and the linearized continuity equation (Eq. \ref{Eq:Linearized_Continuity_Equation}) to
\begin{equation}\label{Eq:Fourier_Continuity_Equation}
    -i\omega\tilde{n_s} + i\vec{k}n_{0,s} \cdot \tilde{v_s} + i\vec{k} \cdot \vec{v_{0,s}}\tilde{n_s} = 0.
\end{equation}

\indent These can then be rearranged to find the values of the new, perturbed quantities (rearranging (Eq. \ref{Eq:Fourier_Momenta_Equation}) and (Eq. \ref{Eq:Fourier_Continuity_Equation}) respectively) in terms of initial quantities
\begin{equation}\label{Eq:Tilde_V_s_Equation}
    \tilde{v_s} = \frac{q_s \tilde{E}}{m_s(\omega - \vec{k} \cdot \vec{v_{0,s}})},
\end{equation}
\begin{equation}\label{Eq:Tilde_n_s_Equation_unsimplified}
    \tilde{n_s} = \frac{n_{0,s}\vec{k} \cdot \vec{v_{0,s}}}{\omega - \vec{k} \cdot \vec{v_{0,s}}},
\end{equation}
which can subsequently be expanded using (Eq. \ref{Eq:Tilde_V_s_Equation})
\begin{equation}\label{Eq:Tilde_n_s_Equation}
    \tilde{n_s} = \frac{q_s n_{0,s}\vec{k}\cdot\tilde{E}}{m_s(\omega - \vec{k}\cdot\vec{v_{0,s}})^2}.
\end{equation}
Taking (Eq. \ref{Eq:Tilde_n_s_Equation}) and substituting it into (Eq. \ref{Eq:Linearized_Gauss_Equation}) recovers
\begin{equation}\label{Eq:Tilde_into_Gauss_Direct}
    i\vec{k} \cdot \tilde{E} = 4\pi \sum_{s} \frac{q_s^2 n_{0,s} \vec{k} \cdot \tilde{E}}{m_s(\omega - \vec{k}\cdot\vec{v_{0,s}})^2}.
\end{equation}
Algebraic shuffling recovers
\begin{equation}\label{Eq:Unsimplified_DR_0}
    i\vec{k} \cdot \tilde{E} - \sum_s \frac{4 \ pi q_s^2 n_{0,s} \vec{k} \cdot \tilde{E}}{m_s(\omega - \vec{k} \cdot \tilde{E})^2} = 0.
\end{equation}

\indent From here, the above expression can be simplified further by replacing terms with $\omega_{p,s}$, that combined with some algebra recovers
\begin{equation}\label{Eq:DR_wp_parenthesis}
    \vec{k} \cdot \tilde{E}\left(i - \sum_s \frac{\omega_{p,s}^2}{(\omega - \vec{k} \cdot \vec{v_{0,s}})}\right) = 0.
\end{equation}
From this equation from, it is easy to see that the above can be cast into $\nabla \cdot \tilde{D} = i\vec{k} \cdot \epsilon\tilde{E} = i\epsilon\vec{k} \cdot \tilde{E}$. Making the parenthetical expression in (Eq. \ref{Eq:DR_wp_parenthesis}) a dielectric constant divided by $i$. Setting this constant to zero recovers a dispersion relation for the system
\begin{equation}\label{Eq:DR_set_to_Zero}
    \frac{1}{i}\left(i - \sum_s \frac{\omega_{p,s}^2}{(\omega - \vec{k} \cdot \tilde{E})^2}\right) = 0.
\end{equation}
Some simplification recovers
\begin{equation}\label{Eq:DR_Species_Form_wk}
    1 = - \sum_s\frac{i\omega_{p,s}^2}{(\omega - \vec{k} \cdot \tilde{E})^2}.
\end{equation}
Recalling now the two species in question ($e^\pm$) we can invoke $\omega_{p,e^-} = \omega_{p,e^+}$, simplifying the expression further. We can also make the assumption that because both species are mixed together in each beam (as opposed to a beam of just $e^-$ and a beam of just $e^+$), we can allow $v_{0,e^-} = v_{0,e^+}$ (no net current). This simplifies (Eq. \ref{Eq:DR_Species_Form_wk}) to
\begin{equation}
    1 + \frac{2i\omega_p^2}{(\omega - \vec{k}\cdot\vec{v_0})^2} = 0,
\end{equation}
where the factor of $2$ comes from the sum of both species. Evaluating the dot product
\begin{equation}\label{Eq:k_dot_V0}
    \vec{k} \cdot \vec{v_0} = k_xv_{0,x} + k_yv_{0,y},
\end{equation}
and recalling that our single dimensional beams have no $y$ velocity ($v_{0,y} = 0$), the final form of the dispersion relation
\begin{equation}\label{Eq:Final_DR}
    1 + \omega_p^2\left(\frac{2i}{(\omega - k_xv_{0,x})^2}\right) = 0,
\end{equation}
with roots
\begin{equation}\label{Eq:Final_DR_Roots}
    \omega(k) = k_xV_0 \pm \omega_p(1 - i).
\end{equation}

\subsubsection{Stability Analysis}
\indent The roots shown in (Eq. \ref{Eq:Final_DR_Roots}) can be recast in the form of 
\begin{equation}\label{Eq:Root_Reform}
    \omega = \omega_R + i\gamma,
\end{equation}
recovering
\begin{equation}\label{Eq:Final_Root_Recast}
    \omega = (k_xV_0 \pm \omega_p) \mp i\omega_p.
\end{equation}
If the roots of the dispersion relation contain only real values ($I(\omega(k)) = 0$), then there is no wave growth or damping. The dispersion relation solutions represent all possible modes. If $I(\omega(k)) \neq 0$, then the electrostatic wave within the system (Eq. \ref{Eq:Plane_Wave_Perturbation_Equation})
can be expressed as
\begin{equation}\label{Eq:E_Static_Wave_DR_Root}
    \tilde{E} = \vec{E_0}e^{i(\vec{k}\cdot\vec{r} - \Re(\omega(k)) t)}e^{\gamma t}.
\end{equation}
Growth rate $\gamma$ greatly influences the time dynamics of the wave. If the parameter is less than zero, the waves are exponentially damped. For greater than zeros, the system has waves that grow exponential and are unstable. \\

\indent For the TSI at the beginning stages of our simulation system ($\vec{B} = 0$), the electrostatic waves generated grow exponentially in the $k_x$ direction at a rate proportional to $\omega_p$ for all $k_x$ values, showing an isotropic wave propagation. Later publications will explore the second occurrence of the TSI after filament mergers where the $B_z$ is decreasing rapidly.

\section{Analysis of Simulation Data}
\subsection{Simulation Set Up}
\indent The analysis is done by running state of the art Particle-in-Cell (PIC) simulations (TRIATAN-MP \cite{TMP}) of two counter streaming electron-positron beams. In particular, four separate parametrizations were simulated. The fiducial simulation had a beam propagation Lorentz factor of $3$, a particle density of $64$ particles per cell, and an electron skin depth of $32$ cells. This simulation is labeled $S1$. To test the convergence of simulation results and study the effect of different parameters on the system, the three additional simulations were performed as follows in reference to the fiducial simulation: halving the Lorentz factor of the beams ($S2$), doubling the particle per cell density ($S3$), or dividing the skin depth by $8$ ($S4$). A full table of simulation parameters can be found in Appendix C.\\

\indent The data from each simulation is then split into $15$ epochs, with each epoch containing $256$ snap shots. This snap shot number per epoch is a requirement of the fast Fourier transform (FFT) spectral analysis and the mirror expansion performed. To ensure the signal is periodic, the mirror of the field is taken along the time axis, creating an array ($2 \times t, x, y)$ that is transformed into ($\omega, k_x, k_y$) with the signal in the $omega$ repeating after $\omega_{max}/2$. The physical data either then plotted directly and studied using energy evolution or is transformed into spectral data, which is used to plot the full dispersion relation and the dispersion relation expansion. The full dispersion relation is a plot of $k$ vs. $\omega/\omega_p$ for the $Log(Amplitude)$ of the FFT field analyzed while the expansion is a $k_x$ vs. $k_y$ plot for each $\omega/\omega_p$ value in the system. 

\subsection{Spectral Cones}
\indent Deducing the varying angular ($\theta$) and frequency ($\omega/\omega_p$) dependencies of the different wave modes and background plasma in each epoch can source a wealth of data about mode behaviors and evolution. To explore these relationships, the method of spectral cone analysis was developed and implemented. This is used to discretize the space to a single angular viewing range with different amplitude limits. This allows a more specific study of salient features. \\

\indent The process of building the spectral cones begins with the component fields. From the magnetic fields studied ($B_z$), the amplitude of the isotropic electromagnetic wave is noted, and the indices within that wave filtered out. These indices correspond across all data fields for the specific field being studied. The indices of the EM wave are then used in the calculations of the spectral cones. The component fields ($E_x$, $E_y$, $B_z$) are first normalized to the total kinetic energy of the system in its initial snapshot
\begin{equation}
    F_i^0 = \frac{F_i}{KE_{Init}}.
\end{equation}
They are then transformed using FFT into the spectral space
\begin{equation}
    \tilde{F_z^0} = FFT(F_z^0).
\end{equation}
The spectral power is defined as follows
\begin{equation}
    \tilde{P_x} = \tilde{E_y^0} * \tilde{B_z^0}; \ \tilde{P_y} = \tilde{-E_x^0} * \tilde{B_z^0},
\end{equation}
\begin{equation}
    ||\tilde{P}|| = \sqrt{\tilde{P_x}^2 + \tilde{P_y}^2}.
\end{equation}
\\
\begin{figure}[h]
    \centering
    \includegraphics[scale=0.2]{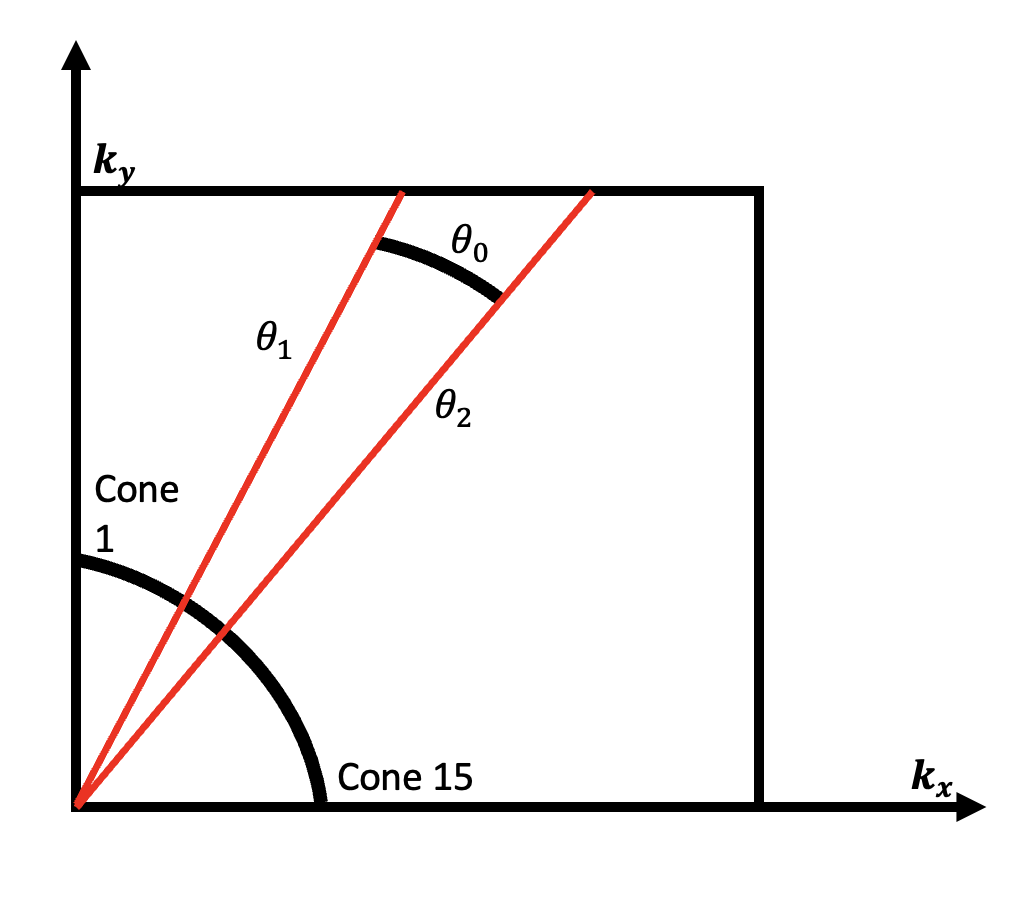}
    \caption{Illustration of a "cone'' in the $k_x, k_y$ box. The box itself represents $\pi/2$ radians.}
    \label{Fig:Cone_Graphic}
\end{figure}
\indent The spectral cone itself is two lines projected onto the $k_xk_y$ space for each omega frequency space value (Fig. \ref{Fig:Cone_Graphic}). Each varying cone sweeps down from the the vertical ($k_y$ or beam perpendicular) to horizontal ($k_x$ or beam parallel) with a constant width of $\frac{\pi}{30}$ radians or $6^{\circ}$ ($15$ cones per $\omega/\omega_p$).
\begin{equation}
    \theta_0 = \frac{\pi/2}{15}.
\end{equation}
The slopes used as cone boundaries are calculated as follows for cone number $C_N$. This is the cone id number $0$ is initial, where the left side of the cone is the y-axis, $14$ correlates to the right side of the cone being the x-axis:
\begin{equation}
    \theta_1^0 = \frac{\pi}{2}; \ \theta_2^0 = \frac{\pi}{2} - \theta_0,
\end{equation}
\begin{equation}
    \theta_1 = \theta_1^0 - (C_N * \theta_0); \ \theta_2 = \theta_2^0 - (C_N * \theta_0),
\end{equation}
\begin{equation}
    slope_1 = tan(\theta_1); \ slope_2 = tan(\theta_2).\\
\end{equation}
\indent The limits of the cones will be displayed in Appendix \ref{App:Spectral_Cone_Limits} for future reference. With the values between the slopes (within the cone) identified, further sorting to discriminate between the excited, high amplitude EM wave and the plasma background follows. The EM wave is found by using amplitude limits taken from the total dispersion relation for the $B_z$ field for the corresponding epoch. Signals within the amplitude limits are those within the EM wave, and are designated \textit{Wave} in the analysis. These are said to be ``wave-like perturbations.'' Those signals outside amplitude bounds of the EM wave are designated \textit{Not Wave} or ``non-wave-like.'' The values within each of the bins are collected for each unique $\omega/\omega_p$ frequency value and then averaged to recover an average spectral power per unique $\omega/\omega_p$ frequency relation. Please see Paper I \cite{MCS_WP1} for more information on total dispersion relations. \\

\section{Results}
\subsection{Total Power Comparisons}
\indent We first look at the cones from a wide frame of reference, where we do not discriminate by signal frequency. The total spectral power of the field amplitudes inside the isotropic wave (wave-like) are summed over each cone and plotted as a function of epoch (time). This is done for all three fields examined ($B_z$, $E_x$, $E_y$) and the same procedure is done for the amplitudes outside of the wave (non-wave like) for comparison. The discussion will begin with $S1$ as the control data set. Then the subsequent data sets and fields will be compared based on their parametric differences. All subsequent discussions will make use of epoch number and characteristic epoch identifiers. These characteristic epochs can be found in Appendix \ref{App:Characteristic_Epochs} for reference.\\

\begin{figure}[h]
    \centering
    \includegraphics[scale=0.3]{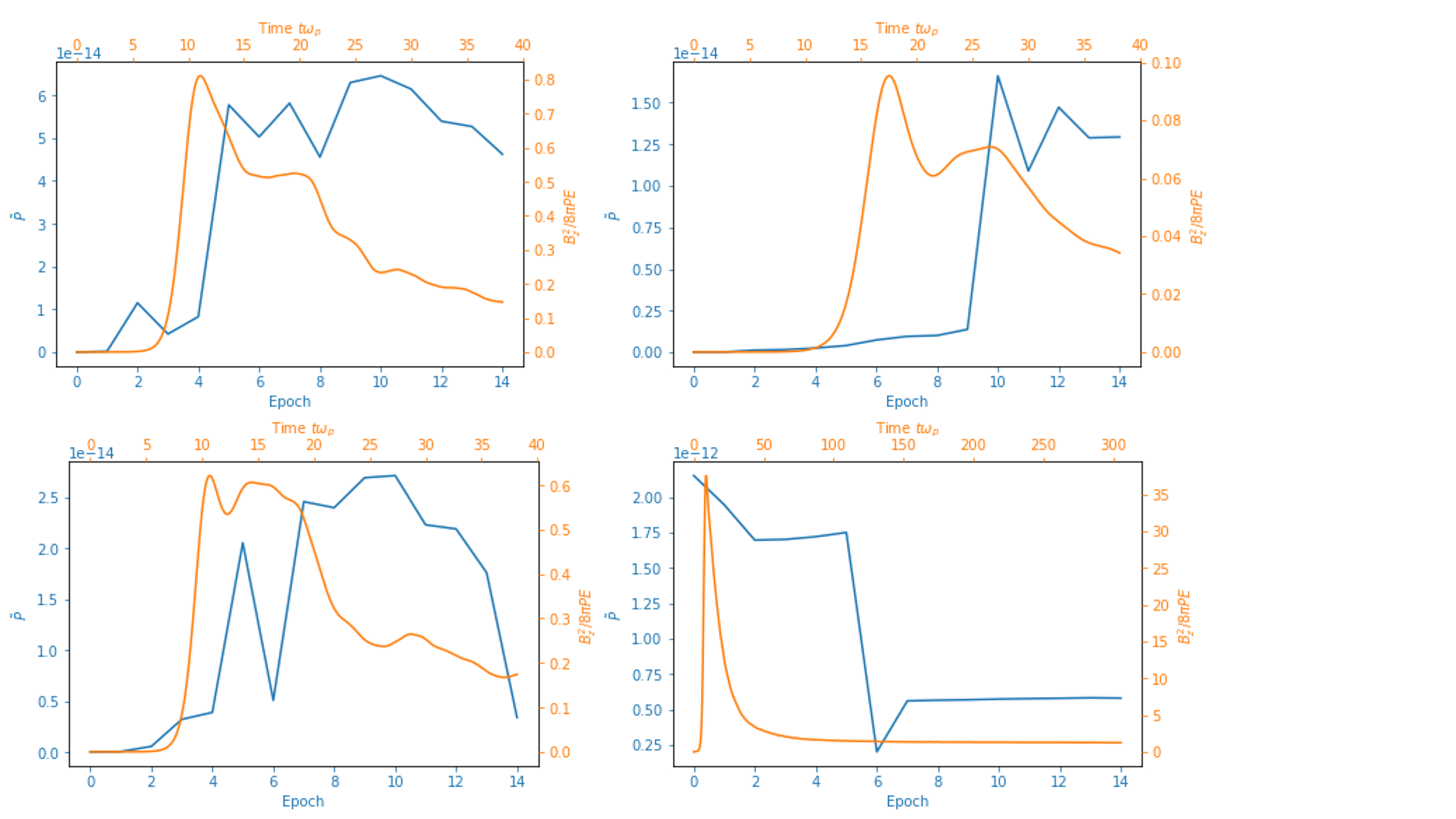}
    \caption{A dual plotting of the total wave power $P_{Tot}(\omega_p t) = \int P(\omega) d\omega$ of both non- and wave-like perturbations of the wave amplitudes (blue) and the magnetic energy density (orange) of the simulations. The major events and characteristic times scales are visible in each panel. Each panel is a different simulation data set with the top left being the fiducial simulation (S$1$), top right the lower beam speed (S$2$), bottom left the higher density (S$3$), and bottom right the low skin depth (S$4$).}
    \label{Fig:Dual_Bz_n_Spectra_Evolutions}
\end{figure}

\indent To begin, it is important to compare the total wave power, $P_{Tot}(\omega_p t) = \int P(\omega) d\omega$, evolution to the physical field evolution ($B_z$) (Fig. \ref{Fig:Dual_Bz_n_Spectra_Evolutions}). These plots illustrate the symbiotic-like relationship between physical fields and particle movements with the spectral waves and radiated power. It can be seen that the total spectral power (both non- and wave-like perturbations) does not neatly follow the same parameter-based behavior relationships as the physical fields do, but what they do share in common is a wave-particle interaction event (for $S1$, $S2$, and $S3$) just before filament merging. Non-wave like refers to anything that is not the isotropic electromagnetic wave. For each of the first three simulations, the total power of non- and wave-like perturbations reaches a local maxima just before the magnetic field peaks post saturation. This is then followed by a local minima as spectral energy is radiated away.  Then, as the physical field falls away, the spectral power continues to grow, possibly indicative of another instability system. With this interaction in mind, the cones can be analyzed in depth, starting from a wide point of view and zooming in all the way to single cone behavior.\\

\begin{figure}[h!]
    \centering
    \includegraphics[scale=0.45]{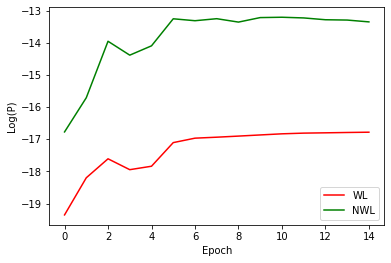}
    \caption{All of the non-wave-like perturbation (NWL - Green) and wave-like perturbation (WL - Red) spectral signals in cones for the magnetic field for each $\omega$ frequency in each epoch are summed and plotted for $S1$ for $B_z$. While small behaviors cannot be seen from this wide view, the overall behavior of the system can be viewed as a whole and various relationships derived, much like the physical field power evolution.}
    \label{Fig:S1_Total_Power_Comp_Bz}
\end{figure}

\begin{figure}[h!]
    \centering
    \includegraphics[scale=0.45]{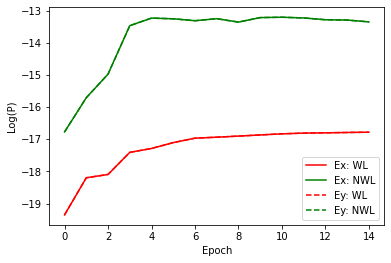}
    \caption{All of the non-wave-like perturbation (NWL - Green) and wave-like perturbation (WL - Red) spectral signals in cones for $E_x$ (solid) and $E_y$ (dashed) for each $\omega$ frequency in each epoch are summed and plotted for $S1$. These fields overlap, displaying only two trend lines. The control simulation sees all three component fields follow the same basic trends as each other.}
    \label{Fig:S1_Total_Power_Comp_ExEy}
\end{figure}

\indent $S1$ shows a rapid increase in magnetic field strength in between system initialization (epoch $1$) and filament ignition (epoch $3$) where it begins to dip until saturation (epoch $5$) (Fig. \ref{Fig:S1_Total_Power_Comp_Bz}). Here, it can be seen that the spectral energy encounters a local minima at the point of filament saturation, where particles are no longer being deflected into nodes. At the filament merger (epoch $8$) spectral power peak is almost indiscernible from the background signals. $E_x$ and $E_y$ show extremely similar spectral signatures compared to each other (Fig. \ref{Fig:S1_Total_Power_Comp_ExEy}). The spectral energy exponentially increases before a plateau at filament saturation, this is consistent with the WI. Each perturbation gradually increases after filament saturation until the simulation ends. In contrast to the local dip at saturation seen in the magnetic field, the electric fields continue their spectral power growth. This behavior at this period of the simulation contrasting with the magnetic field is consistent with the electrostatic TSI, but more discrete trends must be studied.\\

\begin{figure}[h!]
    \centering
    \includegraphics[scale=0.5]{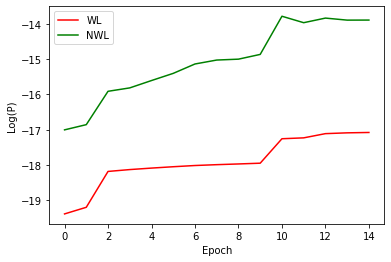}
    \caption{All of the non-wave-like perturbation (NWL - Green) and wave-like perturbation (WL - Red) spectral signals in cones for the magnetic field for each $\omega$ frequency in each epoch are summed and plotted for $S2$. The signal in the low beam speed shows evidence of the signal trending the same as $S1$, but slower (later) in the simulation, while also being smoother. The peak representing filament merger is comparable to $S1$ with the background spectral power being weaker than $S1$ due to beam velocity. The signal corresponding to filament merger is also much later in the simulation (epoch 9/10).}
    \label{Fig:S2_Total_Power_Comp_Bz}
\end{figure}

\begin{figure}[h!]
    \centering
    \includegraphics[scale=0.5]{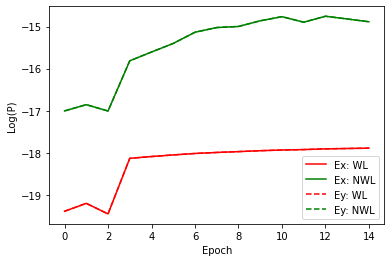}
    \caption{All of the non-wave-like perturbation (NWL - Green) and wave-like perturbation (WL - Red) spectral signals in cones for $E_x$ (solid) and $E_y$ (dashed) for each $\omega$ frequency in each epoch are summed and plotted for $S2$. Similar to the $B_z$ field, the trend is slower and smoother than those of $S1$. We again see the effect of the filament merger with respect to the ambient signals present near filament merging. }
    \label{Fig:S2_Total_Power_Comp_ExEy}
\end{figure}

\indent $S2$ shows a different trend shape in the magnetic field (Fig. \ref{Fig:S2_Total_Power_Comp_Bz}) as $S1$ as the peak begins at epoch $9$ between saturation and filament merger, for which $S2$ displays a longer transitional period accounting for the slower beam speed. The peak for $S2$ becomes more visible due to the power of the filament merger (comparable to $S1$) with respect to the background radiative sources (weaker than $S1$). The power of the non-wave-like perturbation amplitudes is a magnitude lower than that of $S1$ post filament saturation, but the amplitudes for wave-like perturbations are the same. This holds true for the $E_x$ and $E_y$  non-wave-like perturbations (Fig. \ref{Fig:S2_Total_Power_Comp_ExEy}). The wave-like perturbations show a two magnitude decrease in their evolution - consistent with the decreased beam velocity. The trends of the electric fields are different than the magnetic field within $S2$. The non-wave-like amplitudes for both electric fields show a drastic increase between simulation initialization and before filament ignition (epoch $5$) followed by a gradual increase for the remainder of the simulation. Wave-like perturbations gradually increase until a spike and oscillatory behavior just before filament merger (epoch $11$). $E_y$ mimics this trend with minor differences. Overall, the trends for $S2$ and smoother (slower beam velocity means less energy) and later (slower beam velocity means slower evolution) with peaks in the magnetic field amplitudes more visible due to the energy in the merger event.\\

\begin{figure}[h]
    \centering
    \includegraphics[scale=0.45]{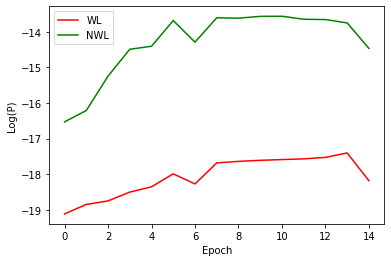}
    \caption{All of the non-wave-like perturbation (NWL - Green) and wave-like perturbation (WL - Red) spectral signals in cones for the magnetic field for each $\omega$ frequency in each epoch are summed and plotted for $S3$. While being on the same time scale of $S1$, $S3$ shows move drastic transitions due to the increased power within the filaments.}
    \label{Fig:S3_Total_Power_Comp_Bz}
\end{figure}

\begin{figure}[h]
    \centering
    \includegraphics[scale=0.45]{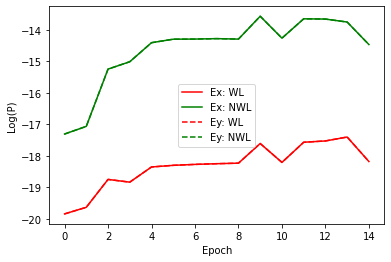}
    \caption{All of the non-wave-like perturbation (NWL - Green) and wave-like perturbation (WL - Red) spectral signals in cones for $E_x$ (solid) and $E_y$ (dashed) for each $\omega$ frequency in each epoch are summed and plotted for $S3$. $S3$ displays drastic differences to $S1$, with more drastic activity (and more activity overall) being seen in the signatures.}
    \label{Fig:S3_Total_Power_Comp_ExEy}
\end{figure}

\indent $S3$ shows similar total spectral power (both non- and wave-like perturbations) magnitudes across all fields as $S1$ (Figs. \ref{Fig:S3_Total_Power_Comp_Bz} and \ref{Fig:S3_Total_Power_Comp_ExEy}), which is to be expected. The peak corresponding to the filament merger is now more discernible from the background signals, as the higher particle density now appears with a larger magnitude signal. The perturbation amplitudes in the magnetic field show a gradual increase until long after merger with a drastic increase at and after filament merger. This non-wave-like perturbation amplitude trend is the same for the electric fields, with the only difference being a power drop at epoch $10$ and peaks at epoch $9$ and $11$ (all of this long after merger at epoch $6$). For the wave itself, the magnetic field trend shows a peak between epochs $4$ and $6$ (saturation at epoch $5$) to a steep drop and gradual increase peaking at epoch $10$. The electric fields follow this but peak at epochs $9$ and $11$ (post merger). The two data sets similar to $S1$ follow the evolutionary trend of WI to filament merger to TSI, with the systematic differences coming from the parameterizations detailed above and in the previous publication. In summation, $S3$ displays more violent transitions of magnitude within the signal trends while maintaining the same time scale of $S1$. This behavior still follows from the physical field implications.\\

\begin{figure}[h]
    \centering
    \includegraphics[scale=0.5]{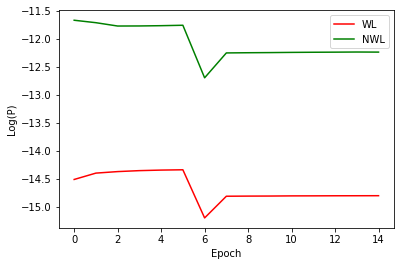}
    \caption{All of the non-wave-like perturbation (NWL - Green) and wave-like perturbation (WL - Red) spectral signals in cones for the magnetic field for each $\omega$ frequency in each epoch are summed and plotted for $S4$. As expected, the signal dies very quickly when there is no filaments to interact with.}
    \label{Fig:S4_Total_Power_Comp_Bz}
\end{figure}

\begin{figure}[h]
    \centering
    \includegraphics[scale=0.5]{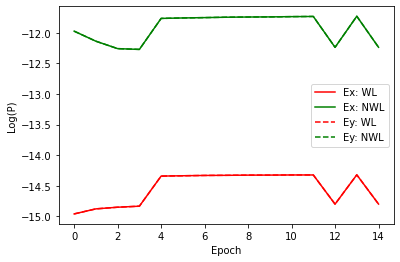}
    \caption{All of the non-wave-like perturbation (NWL - Green) and wave-like perturbation (WL - Red) spectral signals in cones for $E_x$ (solid) and $E_y$ (dashed) for each $\omega$ frequency in each epoch are summed and plotted for $S4$. These signals are plagued by numerical instability once the system is completely evolved and dissipated.}
    \label{Fig:S4_Total_Power_Comp_ExEy}
\end{figure}

\indent $S4$ shows higher total wave power (both non- and wave-like perturbations) magnitudes across all fields (Figs. \ref{Fig:S4_Total_Power_Comp_Bz} and \ref{Fig:S4_Total_Power_Comp_ExEy}). This is expected as there is no large structures or particle-wave interactions to use up power in the system. The non-wave-like perturbation amplitudes show a peak at epoch $5$ to straight line plateauing the rest of the simulation for each field. Wave-like perturbations show some difference between the electric and magnetic fields. The magnetic field shows a decreasing power trend between the initialization of the simulation and epoch $6$, opposite of the other data set's increasing trend. $E_x$ and $E_y$ show a decrease until filament initialization followed by a drastic increase and oscillation. This is explained by the numerical noise in late simulation time. The physical system reaches its end as the small filaments coalesce and smooth out. But, computations continue, evolving the system into a non-physical state. This results in heavy numerical and nonphysical noise that is easily identified as unreal. $S4$ shows evidence of wave particle interaction in the other three simulations by virtue of the absence of a secondary instability\cite{MCS_WP1}. With no filaments to interact with, the EM dissipates and the signal dies. \\

\clearpage
\newpage

\subsection{``Low-Frequency'' vs. ``High-Frequency'' Separation}
\indent Zooming into the data from total power, we now view each simulation using separate ranges of $\omega$ values. These range from low frequency  ($0 \leq \omega < 50$), high frequency ($50 \leq \omega < 100$), and very high frequency ($100 \leq \omega$). Just as before this data is summed over each cone in the box and amplitudes are separated with respect to the dominate electromagnetic wave indices. \\

\begin{figure}[h!]
    \centering
    \begin{subfigure}
        \centering
        \includegraphics[scale=0.5]{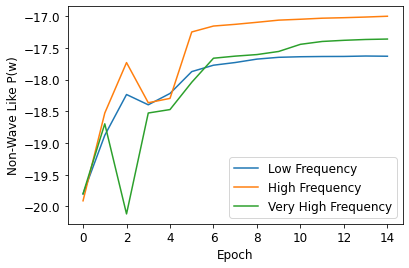}
    \end{subfigure}
    \begin{subfigure} 
        \centering 
        \includegraphics[scale=0.5]{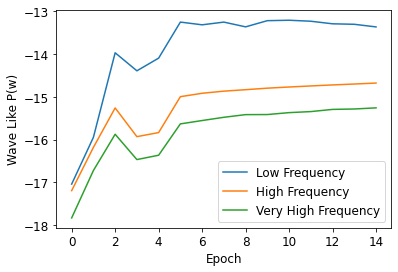}
    \end{subfigure}
    \caption{All of the non-wave-like perturbation (top) and wave-like perturbation (bottom) spectral signals in cones for the magnetic field for each $\omega$ frequency in each epoch are summed and separated by frequency regime and plotted for $S1$. These plots give insights into the different hierarchy of behaviors present in the system. Longer lasting features such as the filaments and their interactions will dominate the low $\omega$. While short lived particle motions occupy the high $\omega$ trends.}
    \label{Fig:S1_Omega_Separation_Bz}
\end{figure}

\begin{figure}[h!]
    \centering
    \begin{subfigure}
        \centering
        \includegraphics[scale=0.25]{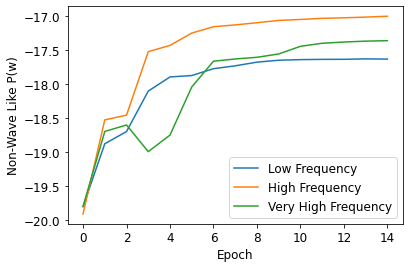}
    \end{subfigure}
    \begin{subfigure} 
        \centering 
        \includegraphics[scale=0.25]{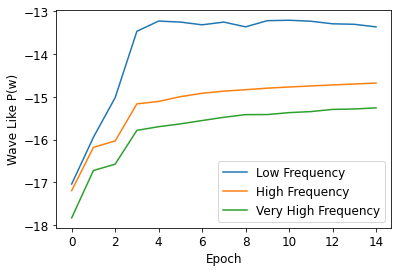}
    \end{subfigure}
    \centering
    \begin{subfigure}
        \centering
        \includegraphics[scale=0.25]{S1_Ex_Summed_External_Power_of_All_Frequencies.png}
    \end{subfigure}
    \begin{subfigure} 
        \centering 
        \includegraphics[scale=0.25]{S1_Ey_Summed_Internal_Power_of_All_Frequencies.png}
    \end{subfigure}
    \caption{All of the non-wave-like perturbation (left) and wave-like perturbation (right) spectral signals in cones for $E_x$ (top) and $E_y$ (bottom) for each $\omega$ frequency in each epoch are summed and separated by frequency regime and plotted for $S1$. With a difference on $\omega$ regimes, we begin to see differences between the component fields of $S1$.}
    \label{Fig:S1_Omega_Separation_ExEy}
\end{figure}

\indent We begin with $S1$, which each other data set will be compared against. In the $B_z$ field (Fig. \ref{Fig:S1_Omega_Separation_Bz}), we see the wave indices span a greater magnitude than those non-wave-like perturbations. For non-wave-like perturbations (top panel), each $\omega$ range plateaus in trend just after saturation, with filament interactions expected in the low $\omega$ trends. The highest $\omega$ frequencies drastically dip right before filament ignition, current filaments begin to emit from particle nodes, and maintain the lowest signal of the regimes until saturation of the filaments. The high and low $\omega$ frequencies both increase before filament ignition and increase again at saturation. Within the isotropic electromagnetic wave (bottom panel), the $\omega$ frequencies remain much more uniform in trend and behavior: increase at filament ignition, dip and increase at saturation, increase and plateau after saturation. For the electric fields (Fig. \ref{Fig:S1_Omega_Separation_ExEy}) we see almost identical behavior to each other. In relation to $B_z$, the magnitude spans are the same and the general behavior seems to follow the magnetic field.\\

\clearpage

\begin{figure}[h!]
    \centering
    \begin{subfigure}
        \centering
        \includegraphics[scale=0.5]{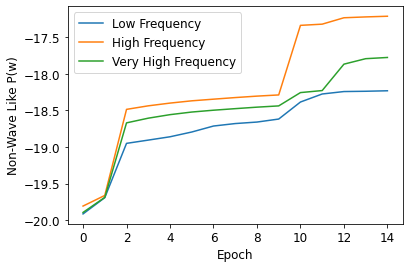}
    \end{subfigure}
    \begin{subfigure} 
        \centering 
        \includegraphics[scale=0.5]{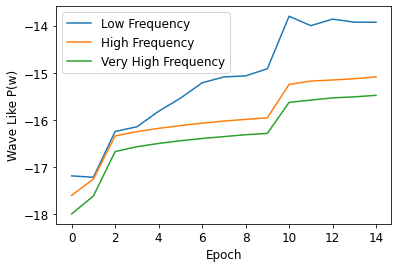}
    \end{subfigure}
    \caption{All of the non-wave-like perturbation (left) and wave-like perturbation (right) spectral signals in cones for the magnetic field for each $\omega$ frequency in each epoch are summed and separated by frequency regime and plotted for $S2$. The slower nature of $S2$ again shows up in the spectral trends, as they are smoother and more delayed than their $S1$ counterpart.}
    \label{Fig:S2_Omega_Separation_Bz}
\end{figure}

\begin{figure}[h!]
    \centering
    \begin{subfigure}
        \centering
        \includegraphics[scale=0.25]{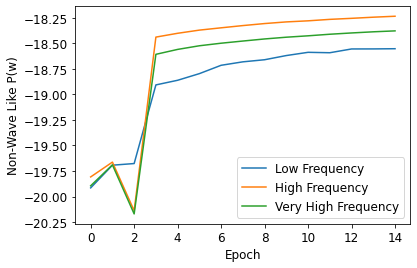}
    \end{subfigure}
    \begin{subfigure} 
        \centering 
        \includegraphics[scale=0.25]{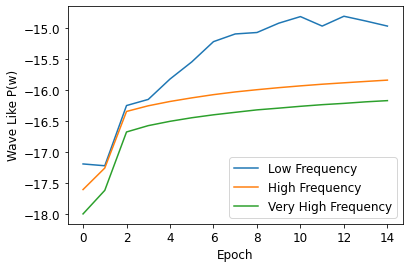}
    \end{subfigure}
    \centering
    \begin{subfigure}
        \centering
        \includegraphics[scale=0.25]{S2_Ex_Summed_External_Power_of_All_Frequencies.png}
    \end{subfigure}
    \begin{subfigure} 
        \centering 
        \includegraphics[scale=0.25]{S2_Ey_Summed_Internal_Power_of_All_Frequencies.png}
    \end{subfigure}
    \caption{All of the non-wave-like perturbation (left) and wave-like perturbation (right) spectral signals in cones for $E_x$ (top) and $E_y$ (bottom) for each $\omega$ frequency in each epoch are summed and separated by frequency regime and plotted for $S2$. $S2$'s electric fields more stratification in its regimes, along with the telltale smoothness.}
    \label{Fig:S2_Omega_Separation_ExEy}
\end{figure}

\indent With $S2$ we can begin to observe changes in spectral behavior based on beam propagation velocity. While magnitude spans are very similar, the trend behavior begins to change immediately. Outside of the wave, all there $\omega$ regimes increase before filament ignition and plateau until before filament merging where they all spike again. Internally, they all spike before filament ignition and we see the low regime not follow the other two for the first time (Fig. \ref{Fig:S2_Omega_Separation_Bz}). While high and very high plateau until just before merging, the low regime steadily increases before overtaking the other two regimes at the same spike location. $E_x$ and $E_y$ again mimic each other closely (Fig. \ref{Fig:S2_Omega_Separation_ExEy}), but much smoother than $S1$. There is a massive spike two epochs before filament ignition where they then plateau for the remainder of the system. Internally to the wave, the behavior is very close to $S1$. Again, $S2$ displays smooth and delayed signals, but the frequency regimes shed more light on new differences.\\

\clearpage

\begin{figure}[h!]
    \centering
    \begin{subfigure}
        \centering
        \includegraphics[scale=0.5]{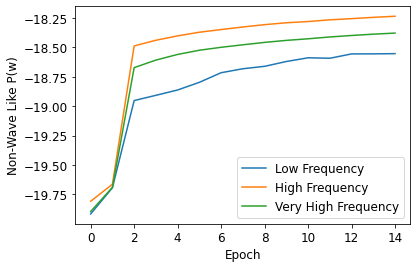}
    \end{subfigure}
    \hfill
    \begin{subfigure} 
        \centering 
        \includegraphics[scale=0.5]{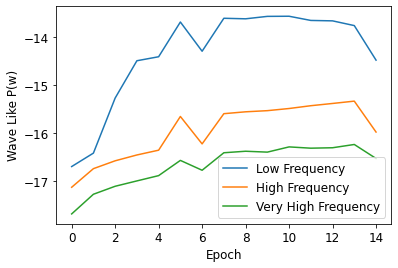}
    \end{subfigure}
    \caption{All of the non-wave-like perturbation (left) and wave-like perturbation (right) spectral signals in cones for the magnetic field for each $\omega$ frequency in each epoch are summed and separated by frequency regime and plotted for $S3$. Due to the higher density filament merger, we see stronger interaction signals. This is evidence of the EM wave interacting with the filament structures.}
    \label{Fig:S3_Omega_Separation_Bz}
\end{figure}

\begin{figure}[h!]
    \centering
    \begin{subfigure}
        \centering
        \includegraphics[scale=0.25]{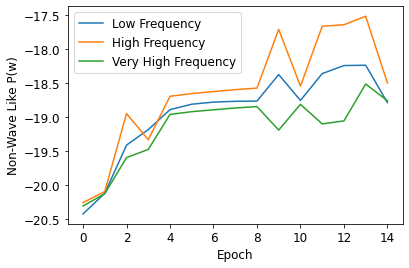}
    \end{subfigure}
    \hfill
    \begin{subfigure} 
        \centering 
        \includegraphics[scale=0.25]{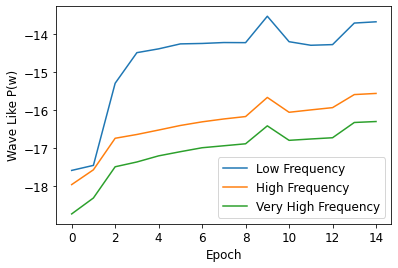}
    \end{subfigure}
    \centering
    \begin{subfigure}
        \centering
        \includegraphics[scale=0.25]{S3_Ex_Summed_External_Power_of_All_Frequencies.png}
    \end{subfigure}
    \hfill
    \begin{subfigure} 
        \centering 
        \includegraphics[scale=0.25]{S3_Ey_Summed_Internal_Power_of_All_Frequencies.png}
    \end{subfigure}
    \caption{All of the non-wave-like perturbation (left) and wave-like perturbation (right) spectral signals in cones for $E_x$ (top) and $E_y$ (bottom) for each $\omega$ frequency in each epoch are summed and separated by frequency regime and plotted for $S3$. Stronger interaction signals are seen in the electric fields due to the higher density (and higher power contained within each filament).}
    \label{Fig:S3_Omega_Separation_ExEy}
\end{figure}

\indent With $S3$ we begin to see changes between the electric fields. The magnetic field sees more complex behavior external to the wave at filament saturation and merger (Fig. \ref{Fig:S3_Omega_Separation_Bz}). While internally we see more movement long after filament merger. $E_x$ displays much more chaotic behavior than before, with spiking before filament ignition and oscillatory behavior long after the filament merger (Fig. \ref{Fig:S3_Omega_Separation_ExEy}). $E_y$ begins to show differences in the low frequency values, with a smoother trend than $E_x$ and the other regimes in $E_y$. The higher density allows for greater spikes near the filament merger, giving evidence for a wave filament interaction that could be the cause of more system excitation.\\

\clearpage

\begin{figure}[h!]
    \centering
    \begin{subfigure}
        \centering
        \includegraphics[scale=0.5]{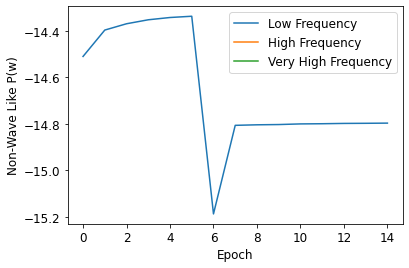}
    \end{subfigure}
    \hfill
    \begin{subfigure} 
        \centering 
        \includegraphics[scale=0.5]{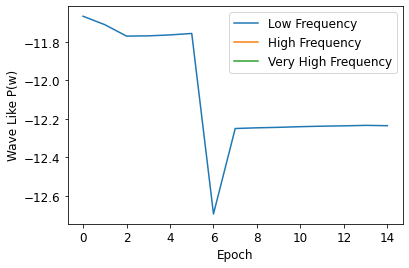}
    \end{subfigure}
    \caption{All of the non-wave-like perturbation (left) and wave-like perturbation (right) spectral signals in cones for the magnetic field for each $\omega$ frequency in each epoch are summed and separated by frequency regime and plotted for $S4$. Without filaments to occupy low $\omega$ space, the only frequencies are generated from particle movement.}
    \label{Fig:S4_Omega_Separation_Bz}
\end{figure}

\begin{figure}[h!]
    \centering
    \begin{subfigure}
        \centering
        \includegraphics[scale=0.25]{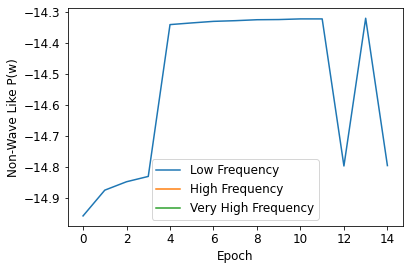}
    \end{subfigure}
    \hfill
    \begin{subfigure} 
        \centering 
        \includegraphics[scale=0.25]{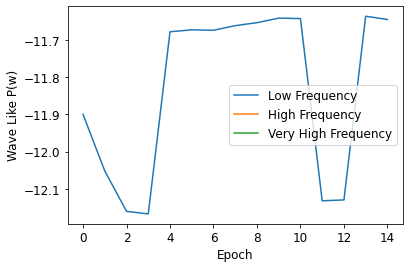}
    \end{subfigure}
    \centering
    \begin{subfigure}
        \centering
        \includegraphics[scale=0.25]{S4_Ex_Summed_External_Power_of_All_Frequencies.png}
    \end{subfigure}
    \hfill
    \begin{subfigure} 
        \centering 
        \includegraphics[scale=0.25]{S4_Ey_Summed_Internal_Power_of_All_Frequencies.png}
    \end{subfigure}
    \caption{All of the non-wave-like perturbation (left) and wave-like perturbation (right) spectral signals in cones for $E_x$ (top) and $E_y$ (bottom) for each $\omega$ frequency in each epoch are summed and separated by frequency regime and plotted for $S4$. These signals are again plagued by numerical instability post system dissipation.}
    \label{Fig:S4_Omega_Separation_ExEy}
\end{figure}

\indent Finally, $S4$'s outlying data is analyzed (Figs. \ref{Fig:S4_Omega_Separation_Bz} and \ref{Fig:S4_Omega_Separation_ExEy}). Despite the lack of substantial physical field evolution, the spectral space shows evolution across the simulation time within all fields. The magnetic field crashes drastically during epoch $5$ and $6$ both inside and outside the wave. The electric fields exhibit chaotic behavior late in the simulation, this can be attributed to the numerical error of progressing a system past the physical boundaries. Without long lasting filaments to occupy the low $\omega$'s, we only see high frequency signals, and more importantly, no interaction signals.\\

\clearpage
\newpage

\subsection{Single Cone Behavior}
\indent Finally, we can view the spectral behavior as a single plot of each cone at each epoch. There are two options to view this massive amount of data: angle as a function of time (cycle through epochs for a single cone then advance cone) or time as a function of angle (cycle through each cone for a single epoch then advance epoch). Both have their advantages and disadvantages when observing patters, below is a summation of both methods condensed into the most important behaviors, trends, and points for different modes seen in each field in each simulation.\\

\subsubsection{S1}

\begin{figure}[h]
    \centering
    \includegraphics[scale=0.5]{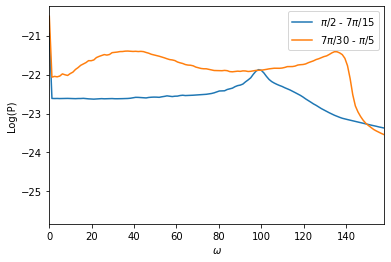}
    \caption{An example of the two modes seen in all for the cones across the simulations. These modes in $S1$ and their motions serve as the comparison basis for $S2$, $S3$, and $S4$. We label the modes as high mode (seen on the right of the signatures) and low mode (seen appearing on the left of the figure).}
    \label{Fig:S1_Single_Cone_Bz_External_Modes}
\end{figure}

\begin{figure}[h]
    \centering
    \begin{subfigure}
        \centering
        \includegraphics[scale=0.25]{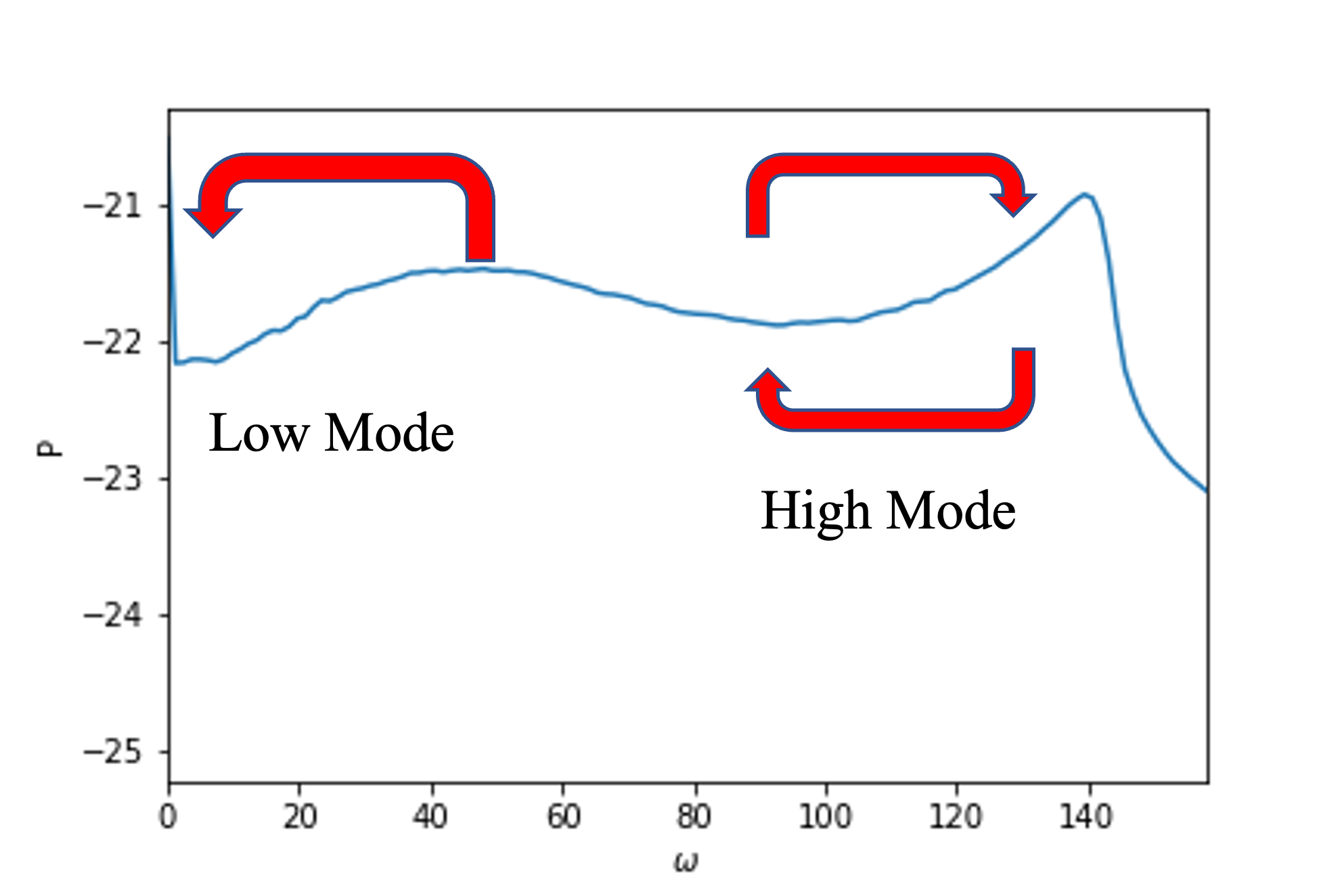}
    \end{subfigure}
    \begin{subfigure} 
        \centering 
        \includegraphics[scale=0.25]{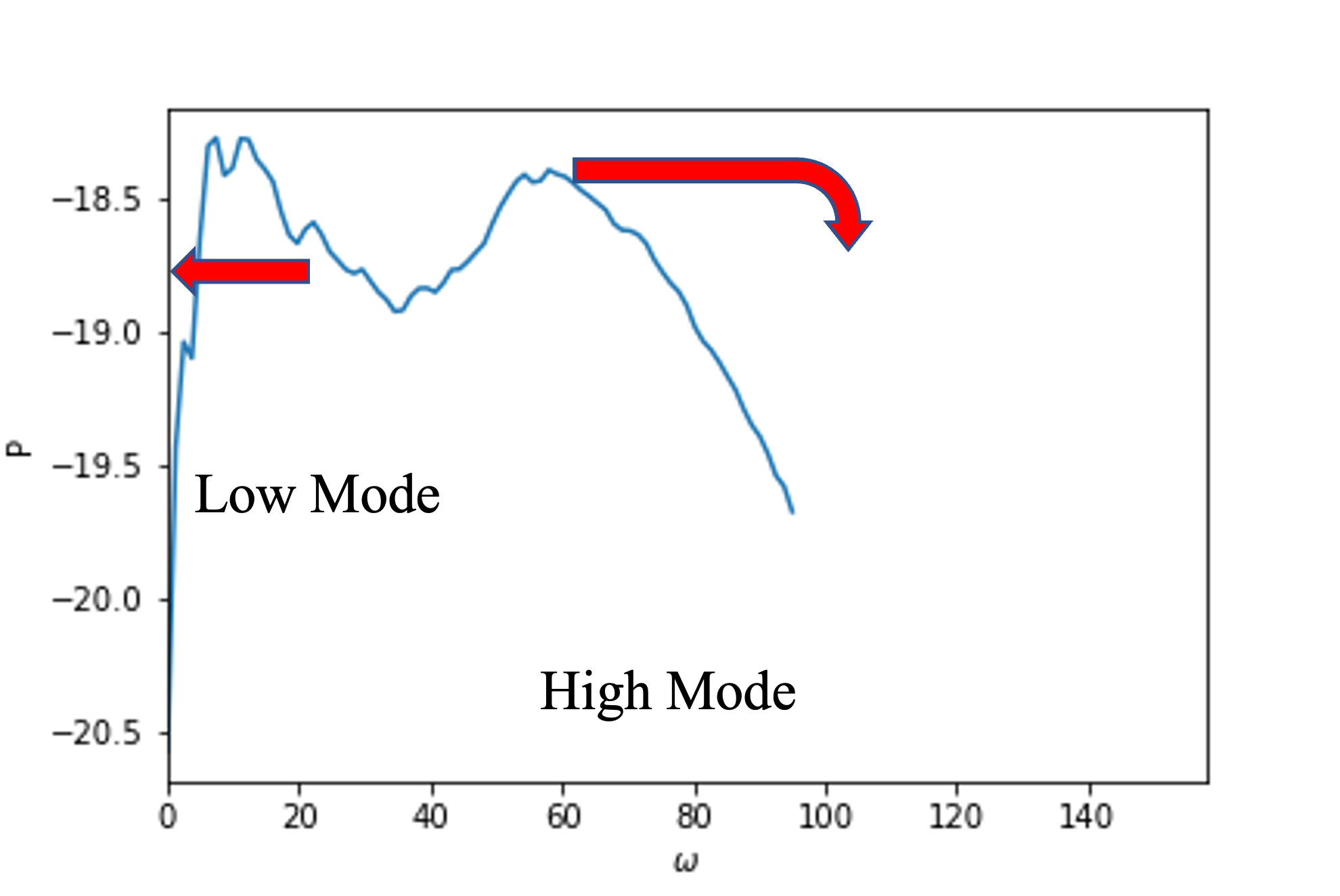}
    \end{subfigure}
    \caption{An example of non-wave-like (left) and wave-like (right) perturbation signals in cones propagating through what is considered the ``normal cycles.'' If the plots from each cone in the epoch (epoch $2$ in this example) are shown in succession, then the modes will follow the red arrows. High mode in the internal cones has a cyclical propagation, reaching its maximum power and frequency at $\frac{\pi}{4}$ and propagating back to its starting point by $\frac{\pi}{2}$. Both low modes propagate left toward $0$, while internal high mode propagates right then down (the tail of the signature makes a whip cracking motion).}
    \label{Fig:Single_Cone_Behavior_Normal_Cycles_Graphic}
\end{figure}

\begin{figure}[h]
    \centering
    \begin{subfigure}
        \centering
        \includegraphics[scale=0.5]{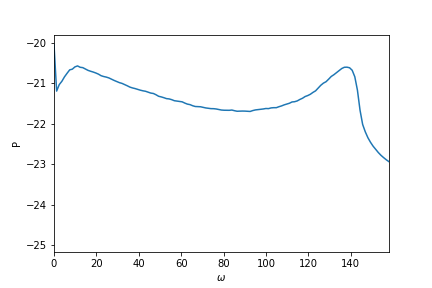}
    \end{subfigure}
    \caption{This signature is found in wave-like perturbations in epoch $3$ (filament ignition) in the angles $\frac{7\pi}{30} - \frac{\pi}{5}$ for $B_z$. Both modes are clearly visible; high mode is in the middle of its propagation cycle while low mode is nearing the end of its propagation cycle.}
    \label{Fig:S1_Single_Cone_Bz_Low_Mode_Shape}
\end{figure}

\indent Beginning with $S1$'s spectral indices outside of the isotropic wave, we immediately see two distinct modes (Fig. \ref{Fig:S1_Single_Cone_Bz_External_Modes}). The first mode (``high mode'') occurs in the higher $\omega$ values. Throughout the simulation, its power increases by three magnitudes as it oscillates from $\omega = 100\omega_p$ to $\omega = 140\omega_p$ and back again by simulations end. This mode is identical in both electric fields. The second mode (``low mode'') appears in the lower $\omega$ values. It first appears at the initialization of the simulation in $\frac{\pi}{6} - \frac{2\pi}{15}$ with frequencies $20\omega_p \leq \omega \leq 80\omega_p$. It appears $\approx \frac{\pi}{2} - \frac{11\pi}{30}$ about $20\omega_p$ lower than the previous epoch as time progresses. We define the cyclical movements of the two modes as a function of cone angle as the ``normal cycles'' (Fig. \ref{Fig:Single_Cone_Behavior_Normal_Cycles_Graphic}). By filament ignition, this low mode has a definite shape at $\omega < 10\omega_p$ but disappears completely by the end of filament ignition (Fig. \ref{Fig:S1_Single_Cone_Bz_Low_Mode_Shape}). The mode reappears for a final time after filament saturation $\frac{\pi}{15} - \frac{\pi}{30}$ as a very small peak in very low $\omega$ values. For the $E_x$, this mode behaves much the same except it reappears after the filament merger as a very weak signal $\approx \frac{\pi}{3} - \frac{\pi}{6}$. $E_y$ sees the same mode shrink drastically from the initialization of the simulation to long after mergers.\\

\begin{figure}[h]
    \centering
    \includegraphics[scale=0.5]{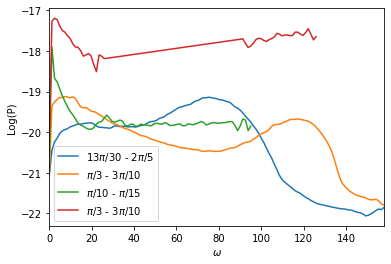}
    \caption{The first (blue) signature is found wave-like perturbations in epoch $1$ (simulation initialization) in the angles $\frac{13\pi}{30} - \frac{2\pi}{5}$ for $B_z$ (all trends). Both high and low modes can be seen for the isotropic EM wave propagating through the system at all times. Next, for the angles $\frac{\pi}{3} - \frac{3\pi}{10}$ (orange), we not only see high mode in its accustomed position, but we see low mode stretch nearly half of the frequency range. Given the early time in the simulation, this can be attributed to the filament building. The green signature is wave-like perturbations found in epoch $2$ (just after simulation initialization) in the angles $\frac{\pi}{10} - \frac{\pi}{15}$. The middle of the $\omega$ range is dominated by a ``broken'' signature - where the trend no longer seems physical. Finally, the red signature is found wave-like perturbations in epoch $3$ (filament ignition) in the angles $\frac{\pi}{3} - \frac{3\pi}{10}$. We see low mode (filaments) receded to its smallest signature and high mode has devolved into a broken signature.}
    \label{Fig:S1_Bz_Modes}
\end{figure}

\indent The indices inside the wave also display two distinct modes. As the simulation is initialized, a mode on the low $\omega$ values (low mode) can immediately be seen and shown to be lower than a second mode with higher $\omega$ values (high mode). Low mode is seen to stretch from $0$ to $80\omega_p$ before shrinking back to only $20\omega_p$. At the same time, high mode began at $60\omega_p$ and propagated right $\approx \frac{\pi}{2} - \frac{2\pi}{15}$ until returning back to $40\omega_p$. Epoch 2 sees the same behavior, but the second mode breaks during its return propagation. A ``broken'' signature refers to a signature or line that looks jagged then completely flat line across the mid region to more jagged and an abrupt ending, looks numerical or nonphysical. During filament ignition, low mode completely disappears by $\frac{\pi}{3} - \frac{3\pi}{10}$, with all lines before it being broken. Meanwhile, low mode has shrunk to only span to $10\omega_p$ (Fig. \ref{Fig:S1_Bz_Modes}). \\

\indent Post filament ignition, the modes regain their normal patterns across the box with high mode no longer propagating leftward $> \frac{7\pi}{30}$. By saturation, low mode has very small return propagation's and low mode continues to span only $10\omega_p$. Once the filamentary structures are saturated and begin drifting to merge, low mode retains its small size and stays relatively stationary while high mode propagates slowly to $100\omega_p$ and its return propagation is a broken signal. For $E_x$, the second mode begins breaking before filament ignition. And for $E_y$, high mode shows a full signal break after filament ignition but before saturation.\\

\clearpage
\newpage 
\subsubsection{S2}
\begin{figure}[h!]
    \centering
    \includegraphics[scale=0.5]{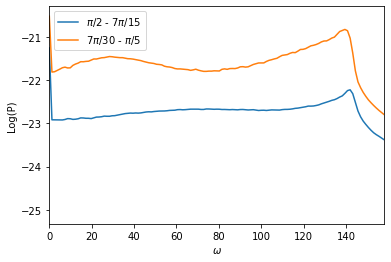}
    \caption{These signatures are found in non-wave-like perturbations in epoch $1$ (blue) and $5$ (orange) in the angles $\frac{\pi}{2} - \frac{7\pi}{15}$ and $\frac{7\pi}{30} - \frac{\pi}{5}$ for $B_z$. We see both low and high mode again, with the difference highlighted here being the shape of the peak of high mode. We see an evolution of the peak shape from more pointed to spread out as the simulation progresses.}
    \label{Fig:S2_Single_Cones_External_Mode_1}
\end{figure}

\begin{figure}[h]
    \centering
    \includegraphics[scale=0.5]{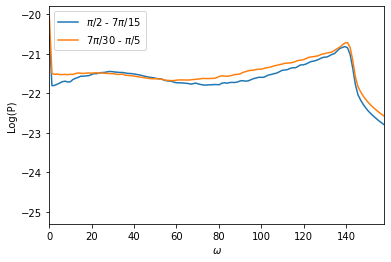}
    \caption{These signatures are found in non-wave-like perturbations in epoch $5$ (blue) and $9$ (orange) in the angles $\frac{\pi}{2} - \frac{7\pi}{15}$ and $\frac{7\pi}{30} - \frac{\pi}{5}$ for $B_z$. Low mode is highlighted here, as the signature begins very shallow compared to $S1$ (indicative by the low beam speed) and as the filaments build the mode begins to take the familiar shape.}
    \label{Fig:S2_Single_Cones_External_Mode_2_Pre_Merge}
\end{figure}

\indent For the lower beam speed run ($S2$), we again begin with the external indices and the study the same two modes seen in $S1$. For this simulation, high mode sees mimicked behaviors with the major difference being the peak shape. The mode's peak is more condensed early in the simulation and widens as time progresses (Fig. \ref{Fig:S2_Single_Cones_External_Mode_1}). Low mode sees more differences in its higher $\omega$ regime. It first appears before filament ignition at $\frac{4\pi}{15} - \frac{7\pi}{30}$ at $20\omega_p$ to $70\omega_p$. The mode remains very shallow in amplitude until just before filament mergers and at $\frac{4\pi}{15} - \frac{2\pi}{15}$ propagates left to $0$ (Fig. \ref{Fig:S2_Single_Cones_External_Mode_2_Pre_Merge}). After merger it becomes more ``peaked'' then ``rounded'' and continues its propagation behavior (Fig. \ref{Fig:S2_Single_Cones_External_Mode_1_Post_Merge}). The electric fields begin to show different behavior than the magnetic counterparts. $E_x$ has the entire signature weaken and flatten before filament ignition begins, epochs $2$ and $3$. The next epoch (directly before ignition) sees low mode appear at $20\omega_p$ and propagate left at angles $\frac{4\pi}{15} - 0$. The mode weakens greatly as the simulation progresses to almost no mode at all (Fig. \ref{Fig:S2_Single_Cones_External_Mode_1_Post_Merge}). In $E_y$, low mode appears just before filament ignition at $40\omega_p$ and while oscillating the entire simulation, becomes a very weak signal in the later half.\\

\begin{figure}[h]
    \centering
    \includegraphics[scale=0.5]{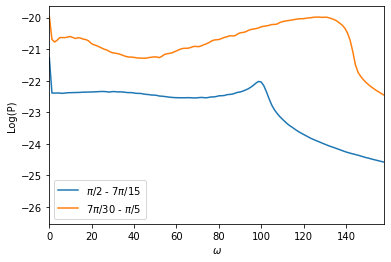}
    \caption{These signatures are found in non-wave-like perturbations in epoch $1$ (blue) and $12$ (orange) in the angles $\frac{\pi}{15} - \frac{\pi}{30}$ and $\frac{4\pi}{15} - \frac{7\pi}{30}$ for $B_z$. Here we see high modes peak again evolve within the simulation, this time post filament merger (wave particle interaction preceding). As its cycle propagates within the epoch, the higher the viewing angle, the more rounded the peak becomes.}
    \label{Fig:S2_Single_Cones_External_Mode_1_Post_Merge}
\end{figure}
\begin{figure}[h]
    \centering
    \includegraphics[scale=0.5]{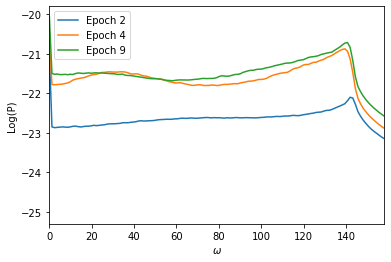}
    \caption{These signatures are found in non-wave-like perturbations in epoch $2$ (blue), $4$ (orange), and $9$ (green) in the angles $\frac{4\pi}{15} - \frac{7\pi}{30}$ (all) for $E_x$. Low mode's evolution from very small signal (pre-filaments) to comparable signal (filamentation and saturation) to barley visible signal (post saturation to pre merger) is shown.}
    \label{Fig:S2_Single_Cones_External_Mode_1_Post_Merge}
\end{figure}

\begin{figure}[h]
    \centering
    \includegraphics[scale=0.5]{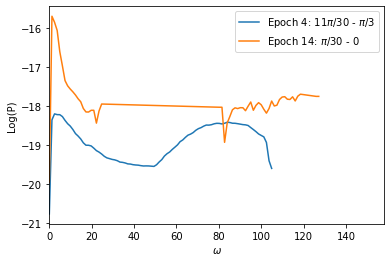}
    \caption{These signatures are found in wave-like perturbations in epoch $4$ (blue) and $14$ (orange) in the angles $\frac{11\pi}{30} - \frac{\pi}{3}$ and $\frac{\pi}{30} - 0$ for $B_z$. In the electric field, we see low mode remain similar to the previous simulations while high mode begins to break down completely.}
    \label{Fig:S2_Single_Cone_Internal_Mode_2_Breakdown}
\end{figure}

\indent Within the electromagnetic wave, the same found in \cite{MCS_WP1}, in $S2$, high mode sees full propagation cycles through just before merger where it then begins to break on the left cycle at $\frac{3\pi}{10} - \frac{7\pi}{30}$ in the magnetic field. Merger and beyond sees high mode downgrade to a very small signal in $\frac{\pi}{2} - \frac{13\pi}{30}$ before breaking completely (Fig. \ref{Fig:S2_Single_Cone_Internal_Mode_2_Breakdown}). High mode stays the same as the non-wave-like perturbation low mode and cycles without deviation. $E_x$ sees high mode more powerful than low mode and breaking right before filament ignition at $\frac{7\pi}{30} - \frac{\pi}{5}$. At ignition, high mode sees a new cycle pattern of extending to higher $\omega$ values, flattening the signature, shrinking, then reappearing $\frac{\pi}{30} - 0$. It no longer breaks from filament ignition onward. $E_y$ sees a second mode more powerful than the one seen in $S1$, more comparable the low mode's peak.\\

\clearpage
\newpage 
\subsubsection{S3}
\begin{figure}[h]
    \centering
    \includegraphics[scale=0.5]{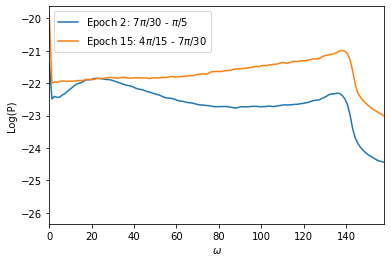}
    \caption{These signatures are found in non-wave-like perturbations in epoch $2$ (blue) and $15$ (orange) in the angles $\frac{7\pi}{30} - \frac{\pi}{5}$ and $\frac{4\pi}{15} - \frac{7\pi}{30}$ for $B_z$. We see low mode's evolution from strong signal to nothing at all (filaments to thermal noise) and high mode's peak evolution once again.}
    \label{Fig:S3_Single_Cone_Mode_1_Evolution}
\end{figure}

\begin{figure}[h]
    \centering
    \includegraphics[scale=0.5]{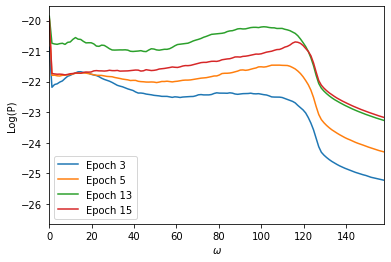}
    \caption{These signatures are found in non-wave-like perturbations in epoch $3$ (blue), $5$ (orange), $13$ (green), and $15$ (red) in the angles $\frac{\pi}{6} - \frac{2\pi}{15}$ (all) for $B_z$. The emergence, disappearance, reemergence, and dissipation of low mode over the simulation period.}
    \label{Fig:S3_Single_Cone_Mode_2_Evolution}
\end{figure}

\begin{figure}[h]
    \centering
    \begin{subfigure}
        \centering
        \includegraphics[scale=0.5]{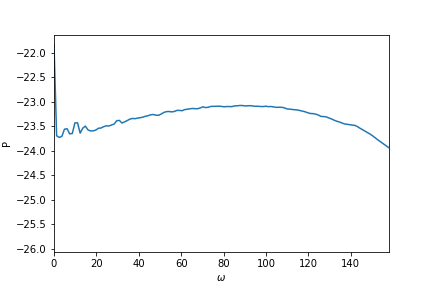}
    \end{subfigure}
    \caption{This signatures are found in non-wave-like perturbations in epoch $1$ in the angles $\frac{7\pi}{30} - \frac{4\pi}{30}$ for $E_y$. This is the first and only time we see the two modes indiscernible in the signal.}
    \label{Fig:S3_Single_Cone_Merged_Mode_Ey}
\end{figure}

\indent $S3$ shows more behavioral changes to the normal cycles of $S1$ than $S2$ to $S1$. External to the wave, high mode begins further out at $80\omega_p$ before returning to normal movement. At saturation, the maxima is becoming thinner and more peaked as the cycle concludes until the end of the simulation where the entire maxima of the mode is a peak (Fig. \ref{Fig:S3_Single_Cone_Mode_1_Evolution}). Low mode is immediately stronger during initialization and appears at $20\omega_p$ to $40\omega_p$ and propagates left toward $0$. During saturation, low mode becomes shallow and is now weaker than high mode and then during merger, exists at $\omega < 20\omega_p$. Post merger, low mode appears and reappears between epochs and fully disappears by the epoch 15 (Fig. \ref{Fig:S3_Single_Cone_Mode_2_Evolution}). $E_x$ sees high mode appear mostly the same. Low mode appears at $\frac{4\pi}{15} - \frac{7\pi}{30}$ and propagates left through $\frac{\pi}{30} - 0$. Each epoch sees the mode weaken until it is barley visible by epochs after filament merging, with only small warbles seen in late epochs. $E_y$ sees both modes appear during the first epoch at $\frac{3\pi}{10} - \frac{4\pi}{15}$ as one continuous curve. It isn't until epoch $2$ that some discernible modes reveal themselves (Fig. \ref{Fig:S3_Single_Cone_Merged_Mode_Ey}). Low mode (low $\omega$) dominates the spectral power until saturation when the high $\omega$ mode becomes more powerful. Normal cycles continue as the low mode becomes more peaked with an odd curved signature appearing in  $\approx \frac{\pi}{3} - \frac{\pi}{6}$. Inside the wave indices in the magnetic field, high mode is a broken signature immediately, and never really recovers a smooth trend, sometimes disappearing completely. Low mode continues to be stable throughout the simulation. $E_x$ and $E_y$ show behavior very close to $S1$. \\


\clearpage
\newpage 
\subsubsection{S4}
\begin{figure}[h]
    \centering
    \includegraphics[scale=0.5]{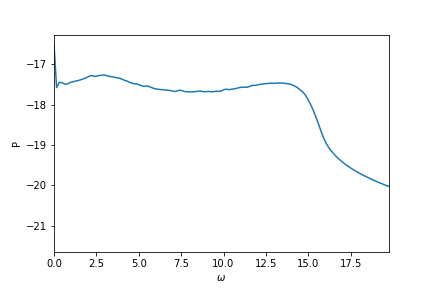}
    \caption{This signature is found in non-wave-like perturbations in epoch $1$ (simulation initialization) in the angles $\frac{2\pi}{15} - \frac{\pi}{10}$ for $B_z$. Only high mode is visible in these spectral signatures, low mode is simply a small bump in the trend, this is due to the N many micro-filaments in the system that cannot fully grow and saturate before smoothing out.}
    \label{Fig:S4_Bz_Normal_Cycle}
\end{figure}

\begin{figure}[h]
    \centering
    \includegraphics[scale=0.5]{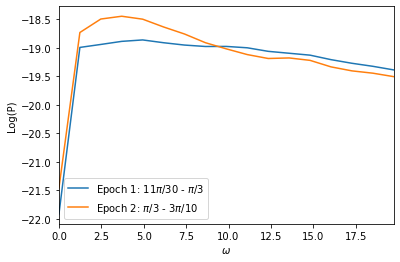}
    \caption{These signatures are found in non-wave-like perturbations in epoch $1$ (blue) and $2$ (orange) in the angles $\frac{11\pi}{30} - \frac{\pi}{3}$ and $\frac{\pi}{3} - \frac{3\pi}{10}$ for $B_z$. As the simulation time increases, the overall magnitude of the signature decreases but when viewing a single epoch's cones in succession, low mode's magnitude increases relative to the rest of the signature.}
    \label{Fig:S4_Bz_Normal_Cycle_Evolution}
\end{figure}

\indent Finally, $S4$ data is viewed. The magnetic field shows only one non-wave-like perturbation mode to the EM wave. Oscillating in a normal cycle (Fig. \ref{Fig:S4_Bz_Normal_Cycle}). Internal to the wave, the magnetic field retains the two mode structure. Low mode increases its magnitude as viewing angle increases but overall magnitude decreases as time increases (Fig. \ref{Fig:S4_Bz_Normal_Cycle_Evolution}). Both electric fields show the same mode behavior across each index regime.\\


\clearpage
\newpage

\section{Conclusion}

\indent In this paper we have shown that Weibel instability system, whose physical parameterization and evolution is described in \cite{MCS_WP1}, has a spectral parameter space equally complex and enlightening. Our previous study showed that there is a coupled electromagnetic wave propagating through the system within each epoch. Total wave power studies in this paper showed that this EM wave interacts with the physical current filaments in the system. This interaction between spectral wave and physical filament, and particles, coincides with the relaxation of Weibel turbulence, occurring just before the highly magnetized and locally unstable environment. These wave-particle interactions before filament merging manifests as energy spikes seen in the $E_x$ field, allowing the electrostatic two stream instability to dominate the pair plasma system by radiating energy away from the EM wave. This radiating energy can be seen in the omega frequency regime study as the wave energy dips and peaks along side the physical energy. As the total wave energy peaks, the magnetic field energy dips, indicating a loss of physical energy.\\

\begin{figure}
    \centering
    \includegraphics[scale=0.5]{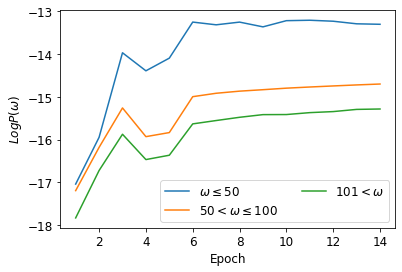}
    \caption{The sum of the power present in cones as a function of $\omega_p$ for each of the labeled frequency regimes. It can be seen that the contribution of the systems power fraction is inversely realted to the frequency magnitude.}
    \label{fig:SumCone_XaxiTime}
\end{figure}

\begin{figure}
    \centering
    \includegraphics[scale=0.5]{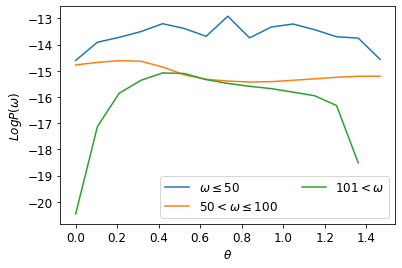}
    \caption{The sum of the power present in cones as a function of $\theta$ for each of the labeled frequency regimes. For the lowest frequency range, a symmetry appears with respect to cone angle. This trend is held to a lesser degree for the remaining ranges.}
    \label{fig:SumCone_XaxiAngle}
\end{figure}

\begin{figure}
        \centering
        \includegraphics[scale=0.5]{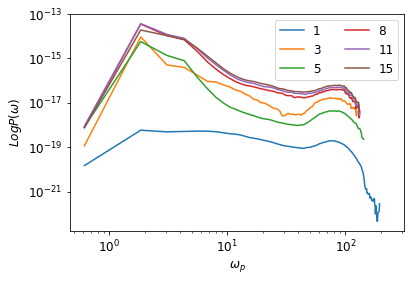}
        \caption{The sum of radiative power over all cones for the characteristic epochs. It is seen that as the WI grows, so does the power. After saturation, the total power stops its major evolution, and more complex behaviors are seen within each epoch.}
        \label{fig:CharEpoch_PowerSum_Freq}
    \end{figure}

\begin{figure}
    \centering
    \includegraphics[scale=0.5]{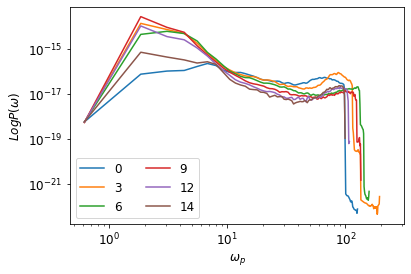}
    \caption{The sum of radiative power over all epochs for an even distribution of cones. Again, the major evolution is seen within the trend lines. The filamentary structure and associated power is seen at lower frequencies where the complex spectral behavior is seen at large frequencies over shorter time scales.}
    \label{fig:CharCones_PowerSum_Freq}
\end{figure}

\indent Analysis of several parameters was done on each of the data sets in an effort to further disentangle the system's dependence of various physical values. Splitting the frequency of radiative power into three regimes, it is seen in (Fig. \ref{fig:SumCone_XaxiTime}) that the low frequency signals radiate the most power within the system. This is in agreement with theory. It is also seen that the largest frequencies contribute the least power. When viewed as a function of cone angle (Fig. \ref{fig:SumCone_XaxiAngle}), symmetry appears in each range, with the most extreme being in the small frequency regime. Finally, the progression of total power is viewed for characteristic epochs (Fig. \ref{fig:CharEpoch_PowerSum_Freq}) and cones (Fig. \ref{fig:CharCones_PowerSum_Freq}). Viewing the characteristic epochs again agrees with theory. As the simulation progresses, the system radiates more power until coming to a stable rate. No true trends are seen in the angles as a function of cone. What is seen are the electromagnetic mode and the electrostatic mode seen in the expanded analysis. \\


\indent The importance of these findings is concrete. After the saturation of the Weibel instability, the violent relaxing of the physical system coincides with the complex spectral behavior of the system and the emergence of the electrostatic TSI. The electromagnetic wave produced by the WI interacts with the current filaments, radiating energy away from the electromagnetic physical system. The energy budget of the entire multi-instability system is now beginning to be filled in. Approximately ten percent is found in the WI. We now see that energy is also contained within the EM wave produced by the physical instability as the system discards its electromagnetic energy in the form of radiation, shown in this analysis. The source of the radiation is thought to be from accelerating particles. The WI plays hosts to many types of radiation, one of which is Jitter radiation. The violent physical relaxation displayed in the previous publication along with the spectral signals shown in this analysis are thought to be new mechanisms for radiation production. Finally, these spectral signals can point to new solutions to the injection problem in collisionless Weibel-mediated shocks. Mergers and wave-particle interactions could serve as acceleration mechanisms for particles to be active participants in diffusive shock acceleration by way of breaching thermal velocities. These questions and theories are currently being investigated using particle data and will be presented in a subsequent publication.\\

\clearpage
\newpage

%\label{}
%\subsection{}
%\subsubsection{}

% If in two-column mode, this environment will change to single-column format so that long equations can be displayed. 
%\begin{widetext}
%$$\mbox{put long equation here}$$
%\end{widetext}

% Figures should be put into the text as floats. 
% Use the graphics or graphicx packages (distributed with LaTeX2e).
% Put the label that you will use with \ref{} command in the braces of the \label{} command.
%
% \begin{figure}
% \end{figure}

% \begin{table}
% \end{tabular}
% \end{table}

\begin{acknowledgments}
This work was supported by NSF grant PHY-2010109. The author would like to thank Dr. Medvedev for his guidance and role of graduate advisor for this work and subsequent works. This research made use of "Tristan-MP v2" particle-in-cell code.
\end{acknowledgments}

\appendix

\section{\label{App:Simulation_Parameters}Simulation Parameters}
\indent Below is a table of parameters for the TRISTAN-MP simulation software corresponding to the four simulations ran in this and subsequent studies. We encourage the reader to review the appendices in
\cite{MCS_WP1} for a full treatment and explanation of the code normalization and their counterparts in cgs. We also encourage the reader to view \cite{26} and \cite{27} for more information. \\

\clearpage 
\onecolumngrid

\begin{table}[h]
\begin{center}
\caption{Parameters for the simulation code TRISTAN-MP.}
\begin{tabular}{ |p{5.0cm}||p{2.5cm}|p{3cm}|p{3cm}|p{3cm}|  }
 \hline
 \multicolumn{5}{|c|}{TRISTAN-MP Parameters} \\
 \hline
Simulation Parameter & S1 - Fiducial & S2 - Low $\gamma$ & S3 - High $PPC$ & S4 - Low $d_{e^-}$ \\
 \hline
 \multicolumn{5}{|c|}{Universal Parameters}\\
  \hline
 X Direction CPUs & \multicolumn{4}{|c|}{16}\\
 Y Direction CPUs & \multicolumn{4}{|c|}{4}\\
 X Direction Grid & \multicolumn{4}{|c|}{512}\\
 Y Direction Grid & \multicolumn{4}{|c|}{512}\\
 c (Comp. Units) & \multicolumn{4}{|c|}{.45}\\
 Correction to c & \multicolumn{4}{|c|}{1.025}\\
 Grid interval & \multicolumn{4}{|c|}{1}\\
 Timestep Interval & \multicolumn{4}{|c|}{1}\\
 Smoothing Filter Passes & \multicolumn{4}{|c|}{0}\\
 Magnetization Number & \multicolumn{4}{|c|}{0.00}\\
 Max Number of Particles & \multicolumn{4}{|c|}{$ 1 \times 10^9$}\\
 Delta $\gamma$: ($\frac{kT_i}{m_ic^2}$) & \multicolumn{4}{|c|}{0.2}\\
 Magnitization Number ($\sigma_0$) & \multicolumn{4}{|c|}{0}\\
 \hline
 \multicolumn{5}{|c|}{Simulation Input Parameters}\\
 \hline
 $\gamma_{Beam}$ & 3 & 1.5 & 3 & 3\\
 Particles per Cell & 64 & 64 & 128 & 64\\
 $e^-$ Skin Depth in Cell & 32 & 32 & 32 & 4\\
 \hline
 \multicolumn{5}{|c|}{Growth Rate}\\
 \hline
 Theoretical Rate ($\Gamma_{WI}^T$) & 58.07 & 82.12 & 58.07 & 7.25\\
 Simulation Rate ($\Gamma_{WI}^S$) & 61.06 & 61.43 & 61.06 & 7.94\\
 %Analysis Normalization ($2\pi\omega_p$) & 631.98 & 631.98 & 631.98 & 78.98\\ %The code that produced the plot does not include the rouge 2\pi. 
 Growth Rate Ratios & 0.95 & 1.33 & 0.95 & 0.91\\
 \hline
 \multicolumn{5}{|c|}{Energy in Comp. Units}\\
 \hline
 Initial KE & 61884625.61 & 30941686.08 & 123767220.92 & 61884624.50 \\
 Initial KE Density & 236.07 & 118.03 & 472.13 & 236.07 \\
 Initial Electric PE & 0.0011 & 0.0003 & 0.0005 & $1.79 x 10^{-5}$ \\
 Initial Electric PE Density & $4.37 \times 10^{-9}$ & $1.45 \times 10^{-9}$ & $2.18 \times 10^{-9}$ & 4.69\\
 \hline
 \multicolumn{5}{|c|}{Magnetic Energy and Time in $\omega_p$}\\
 \hline
 $B_{Max}$ at Saturation & 0.81 & 0.09 & 0.62 & 37.55 \\
 $B_{Max}$ at Merger & 0.52 & 0.07 & 0.60 & N/A \\
 $\omega_p$ at $B_{Max}$ at Saturation & 11.07 & 17.52 & 10.67 & 8.75 \\
 $\omega_p$ at $B_{Max}$ at Merger & 19.49 & 26.56 & 14.40 & N/A \\
 \hline
 \multicolumn{5}{|c|}{Simulation Derived Parameters}\\
 \hline
 $\omega_{p}^{e^\pm}$ (Comp. Units) & 71.12 & 71.12 & 71.12 & 8.88\\
 $\omega_{p}$ (Comp. Units) & 100.58 & 100.58 & 100.58 & 12.57\\
 Skin Depth $d_{e^\pm} = \frac{c}{\omega_{p}^{e^\pm}}$ & 0.006 & 0.006  & 0.006 & 0.05 \\
 Skin Depth $d_{p} = \frac{c}{\omega_{p}^{p}}$ & 0.004 & 0.004 & 0.004 & 0.035 \\
 Step Size ($\omega_p^{-1})$ & 100.58 & 100.58 & 100.58 & 12.57 \\
 Cell Size ($c/\omega_p$) & 0.03125 & 0.03125 & 0.03125 & 0.25 \\
 Plasma React. Time $\frac{c}{d_{e^-} in \ Cells}$ & & & &\\
 \hline
 \multicolumn{5}{|c|}{TRISTAN-MP Wiki Calculations}\\
 \hline
 $\Delta x$ & \multicolumn{4}{|c|}{1 cm}\\
 $\Delta t$ & \multicolumn{4}{|c|}{$1.501 \times 10^{-11} s$}\\
 $n_{e^\pm}$ Species Num. Density & \multicolumn{4}{|c|}{$1 \times 10^{15} s$}\\
 $n_0$ Total Num. Density & \multicolumn{4}{|c|}{$2 \times 10^{15} s$}\\
 $w_0$ $\frac{Real Particles}{Single Macroparticle}$ & $3.125 \times 10^{13}$ & $3.125 \times 10^{13}$ & $1.5625 \times 10^{13}$ & $3.125 \times 10^{13}$\\
 Formula $\omega_{p,s}$ & \multicolumn{4}{|c|}{$1783255450012.7007 s^{-1}$}\\
 Formula $d_s$ &  \multicolumn{4}{|c|}{$2.5234746934141993e-13 cm   $}\\
 Fiducial $\omega_{e^-}^0$ & 0.0140625 & 0.0140625 & 0.0140625 & 0.1125 \\
 Fiducial $d_{e^-}^0$ & 32 & 32 & 32 & 4\\
 Actual $\omega_{p,s}$ & 0.0099 & 0.0099 & 0.0099 & 0.0795\\
 Actual $d_s$ & 45.2548 & 45.2548 & 45.2548 & 5.6568 \\
 \hline
\end{tabular}
\end{center}
\end{table}

\clearpage
\newpage

\section{\label{App:Characteristic_Epochs}Characteristic Epochs}
\begin{table}[h]
\begin{center}
\caption{Characteristic epoch numbers and the corresponding behavior for the four simulation data sets.}
\begin{tabular}{ |p{4.5cm}||p{3cm}|p{3cm}|p{3cm}|p{3cm}|  }
 \hline
 \multicolumn{5}{|c|}{Characteristic Epochs} \\
 \hline
Simulation Parameter & S1 - Fiducial & S2 - Low $\gamma$ & S3 - High $PPC$ & S4 - Low $d_{e^-}$ \\
\hline
\hline
Thermal Noise & 1 & 1 & 1 & 1\\
Filament Ignition & 3 & 5 & 3 & N/A\\
Saturation & 5 & 7 & 5 & N/A\\
Filament Merger & 8 & 11 & 6 & N/A\\
Weibel Dissipation & 11 & 12 & 10 & N/A\\
Thermal/Numerical Noise & 15 & 15 & 15 & 15\\
\hline
\end{tabular}
\end{center}
\end{table}

\section{\label{App:Spectral_Cone_Limits}Spectral Cone Limits}
\begin{table}[h]
\begin{center}
\caption{Spectral cone limits in both radians and angles. The leftmost bounds are closer to the y-axis while the rightmost bounds are those closest to the x-axis}
\begin{tabular}{ |p{4.5cm}||p{3cm}|p{3cm}|p{3cm}|p{3cm}|  }
 \hline
 \multicolumn{5}{|c|}{Cone Limits} \\
 \hline
Cone Identifier & Left Bound Radians & Right Bound Radians & Left Bound Angle & Right Bound Angle \\
\hline
\hline
$0$ & $\pi/2$ & $7\pi/15$ & $90^{\circ}$ & $84^{\circ}$\\
$1$ & $7\pi/15$ & $13\pi/30$ & $84^{\circ}$ & $78^{\circ}$\\
$2$ & $13\pi/30$ & $2\pi/5$ & $78^{\circ}$ & $72^{\circ}$\\
$3$ & $2\pi/5$ & $11\pi/30$ & $72^{\circ}$ & $66^{\circ}$\\
$4$ & $11\pi/30$ & $\pi/3$ & $66^{\circ}$ & $60^{\circ}$\\
$5$ & $\pi/3$ & $3\pi/10$ & $60^{\circ}$ & $54^{\circ}$\\
$6$ & $3\pi/10$ & $4\pi/15$ & $54^{\circ}$ & $48^{\circ}$\\
$7$ & $4\pi/15$ & $7\pi/30$ & $48^{\circ}$ & $42^{\circ}$\\
$8$ & $7\pi/30$ & $\pi/5$ & $42^{\circ}$ & $36^{\circ}$\\
$9$ & $\pi/5$ & $\pi/6$ & $36^{\circ}$ & $30^{\circ}$\\
$10$ & $\pi/6$ & $2\pi/15$ & $30^{\circ}$ & $24^{\circ}$\\
$11$ & $2\pi/15$ & $\pi/10$ & $24^{\circ}$ & $18^{\circ}$\\
$12$ & $\pi/10$ & $\pi/15$ & $18^{\circ}$ & $12^{\circ}$\\
$13$ & $\pi/15$ & $\pi/30$ & $12^{\circ}$ & $6^{\circ}$\\
$14$ & $\pi/30$ & $0$ & $6^{\circ}$ & $0^{\circ}$\\
\hline
\end{tabular}
\end{center}
\end{table}

\twocolumngrid
\nocite{*}
\bibliographystyle{abbrv}
\bibliography{main}

\end{document}